\documentclass[twocolumn]{aa}

\usepackage{graphicx}
\usepackage{amsmath,amsfonts,amssymb}
\usepackage{txfonts}
\usepackage{color}
\usepackage{fixltx2e}
\usepackage{natbib}
\usepackage{caption,subcaption}
\usepackage{epsfig}
\usepackage{float}
\usepackage{afterpage}
\usepackage{ifthen}
\usepackage[morefloats=12]{morefloats}
\usepackage{placeins}
\usepackage{multicol}
\usepackage{siunitx}
\DeclareSIUnit\parsec{pc}
\DeclareSIUnit\lightyear{ly}
\DeclareSIUnit\Kcmb{K_{cmb}}
\DeclareSIUnit\year{yr}
\bibpunct{(}{)}{;}{a}{}{,}
\usepackage[switch]{lineno}
\definecolor{linkcolor}{rgb}{0.6,0,0}
\definecolor{citecolor}{rgb}{0,0,0.75}
\definecolor{urlcolor}{rgb}{0.12,0.46,0.7}
\usepackage[breaklinks, colorlinks, urlcolor=urlcolor,
    linkcolor=linkcolor,citecolor=citecolor,pdfencoding=auto]{hyperref}
\hypersetup{linktocpage}

\def\setsymbol#1#2{\expandafter\def\csname #1\endcsname{#2}}
\def\getsymbol#1{\csname #1\endcsname}

\def\Planck{\textit{Planck}}

\newbox\tablebox    \newdimen\tablewidth
\def\leaderfil{\leaders\hbox to 5pt{\hss.\hss}\hfil}
\def\endPlancktable{\tablewidth=\columnwidth 
    $$\hss\copy\tablebox\hss$$
    \vskip-\lastskip\vskip -2pt}
\def\endPlancktablewide{\tablewidth=\textwidth 
    $$\hss\copy\tablebox\hss$$
    \vskip-\lastskip\vskip -2pt}
\def\tablenote#1 #2\par{\begingroup \parindent=0.8em
    \abovedisplayshortskip=0pt\belowdisplayshortskip=0pt
    \noindent
    $$\hss\vbox{\hsize\tablewidth \hangindent=\parindent \hangafter=1 \noindent
    \hbox to \parindent{$^#1$\hss}\strut#2\strut\par}\hss$$
    \endgroup}
\def\doubleline{\vskip 3pt\hrule \vskip 1.5pt \hrule \vskip 5pt}

\def\L2{\ifmmode L_2\else $L_2$\fi}

\def\DeltaT{\ifmmode \Delta T\else $\Delta T$\fi}
\def\deltat{\ifmmode \Delta t\else $\Delta t$\fi}
\def\fknee{\ifmmode f_{\rm knee}\else $f_{\rm knee}$\fi}
\def\Fmax{\ifmmode F_{\rm max}\else $F_{\rm max}$\fi}
\def\solar{\ifmmode{\rm M}_{\mathord\odot}\else${\rm M}_{\mathord\odot}$\fi}
\def\Msolar{\ifmmode{\rm M}_{\mathord\odot}\else${\rm M}_{\mathord\odot}$\fi}
\def\Lsolar{\ifmmode{\rm L}_{\mathord\odot}\else${\rm L}_{\mathord\odot}$\fi}
\def\inv{\ifmmode^{-1}\else$^{-1}$\fi}
\def\mo{\ifmmode^{-1}\else$^{-1}$\fi}
\def\sup#1{\ifmmode ^{\rm #1}\else $^{\rm #1}$\fi}
\def\expo#1{\ifmmode \times 10^{#1}\else $\times 10^{#1}$\fi}
\def\,{\thinspace}
\def\lsim{\mathrel{\raise .4ex\hbox{\rlap{$<$}\lower 1.2ex\hbox{$\sim$}}}}
\def\gsim{\mathrel{\raise .4ex\hbox{\rlap{$>$}\lower 1.2ex\hbox{$\sim$}}}}

\def\simprop{\mathrel{\raise .4ex\hbox{\rlap{$\propto$}\lower 1.2ex\hbox{$\sim$}}}}
\def\deg{\ifmmode^\circ\else$^\circ$\fi}
\def\pdeg{\ifmmode $\setbox0=\hbox{$^{\circ}$}\rlap{\hskip.11\wd0 .}$^{\circ}
          \else \setbox0=\hbox{$^{\circ}$}\rlap{\hskip.11\wd0 .}$^{\circ}$\fi}
\def\arcs{\ifmmode {^{\scriptstyle\prime\prime}}
          \else $^{\scriptstyle\prime\prime}$\fi}
\def\arcm{\ifmmode {^{\scriptstyle\prime}}
          \else $^{\scriptstyle\prime}$\fi}
\newdimen\sa  \newdimen\sb
\def\parcs{\sa=.07em \sb=.03em
     \ifmmode \hbox{\rlap{.}}^{\scriptstyle\prime\kern -\sb\prime}\hbox{\kern -\sa}
     \else \rlap{.}$^{\scriptstyle\prime\kern -\sb\prime}$\kern -\sa\fi}
\def\parcm{\sa=.08em \sb=.03em
     \ifmmode \hbox{\rlap{.}\kern\sa}^{\scriptstyle\prime}\hbox{\kern-\sb}
     \else \rlap{.}\kern\sa$^{\scriptstyle\prime}$\kern-\sb\fi}
\def\ra[#1 #2 #3.#4]{#1\sup{h}#2\sup{m}#3\sup{s}\llap.#4}
\def\dec[#1 #2 #3.#4]{#1\deg#2\arcm#3\arcs\llap.#4}
\def\deco[#1 #2 #3]{#1\deg#2\arcm#3\arcs}
\def\rra[#1 #2]{#1\sup{h}#2\sup{m}}

\def\dots{\relax\ifmmode \ldots\else $\ldots$\fi}
\def\WHzsr{\ifmmode $W\,Hz\mo\,sr\mo$\else W\,Hz\mo\,sr\mo\fi}
\def\mHz{\ifmmode $\,mHz$\else \,mHz\fi}
\def\GHz{\ifmmode $\,GHz$\else \,GHz\fi}
\def\mKs{\ifmmode $\,mK\,s$^{1/2}\else \,mK\,s$^{1/2}$\fi}
\def\muKs{\ifmmode \,\mu$K\,s$^{1/2}\else \,$\mu$K\,s$^{1/2}$\fi}
\def\muKRJs{\ifmmode \,\mu$K$_{\rm RJ}$\,s$^{1/2}\else \,$\mu$K$_{\rm RJ}$\,s$^{1/2}$\fi}
\def\muKHz{\ifmmode \,\mu$K\,Hz$^{-1/2}\else \,$\mu$K\,Hz$^{-1/2}$\fi}
\def\MJysr{\ifmmode \,$MJy\,sr\mo$\else \,MJy\,sr\mo\fi}
\def\MJysrmK{\ifmmode \,$MJy\,sr\mo$\,mK$_{\rm CMB}\mo\else \,MJy\,sr\mo\,mK$_{\rm CMB}\mo$\fi}
\def\microns{\ifmmode \,\mu$m$\else \,$\mu$m\fi}

\def\muK{\ifmmode \,\mu$K$\else \,$\mu$\hbox{K}\fi}
\def\microK{\ifmmode \,\mu$K$\else \,$\mu$\hbox{K}\fi}
\def\muW{\ifmmode \,\mu$W$\else \,$\mu$\hbox{W}\fi}
\def\kms{\ifmmode $\,km\,s$^{-1}\else \,km\,s$^{-1}$\fi}
\def\kmsMpc{\ifmmode $\,\kms\,Mpc\mo$\else \,\kms\,Mpc\mo\fi}
\providecommand{\sorthelp}[1]{}

\def\WMAP{\textit{WMAP}}

\def\LCDM{$\Lambda$CDM}

\def\commander{\texttt{Commander}}

\renewcommand{\d}[0]{\vec{d}}
\renewcommand{\t}[0]{\vec{t}}
\newcommand{\A}[0]{\tens{A}}

\newcommand{\n}[0]{\vec{n}}
\newcommand{\x}[0]{\vec{x}}

\newcommand{\s}[0]{\vec{s}}
\renewcommand{\a}[0]{\vec{a}}
\newcommand{\m}[0]{\vec{m}}

\newcommand{\B}[0]{\tens{B}}

\newcommand{\Cp}[0]{\tens{C}}
\renewcommand{\L}[0]{\tens{L}}
\newcommand{\g}[0]{\vec{g}}
\newcommand{\N}[0]{\tens{N}}

\newcommand{\M}[0]{\tens{M}}

\renewcommand{\S}[0]{\tens{S}}

\renewcommand{\P}[0]{\tens{P}}

\newcommand{\Dbp}[0]{\Delta_{\mathrm{bp}}}

\newcommand{\BP}{\textsc{BeyondPlanck}}
\newcommand{\lfi}[0]{LFI}
\newcommand{\hfi}[0]{HFI}
\newcommand{\npipe}[0]{\texttt{NPIPE}}

\newcommand{\srollTwo}[0]{\texttt{SROLL2}}

\def\inv{^{-1}}

\begin{document}

\title{\bfseries{\scshape{BeyondPlanck}} XII. Cosmological parameter constraints with end-to-end error propagation}
\newcommand{\oslo}[0]{3}
\newcommand{\milanoA}[0]{1}
\newcommand{\milanoB}[0]{4}
\newcommand{\milanoC}[0]{2}
\newcommand{\triesteB}[0]{5}
\newcommand{\planetek}[0]{6}
\newcommand{\princeton}[0]{7}
\newcommand{\jpl}[0]{8}
\newcommand{\helsinkiA}[0]{9}
\newcommand{\helsinkiB}[0]{10}
\newcommand{\nersc}[0]{11}
\newcommand{\haverford}[0]{12}
\newcommand{\mpa}[0]{13}
\newcommand{\triesteA}[0]{14}
\author{\small
S.~Paradiso\inst{\milanoA, \milanoC}\thanks{Corresponding author: S.~Paradiso; \url{simone.paradiso@unimi.it}}
\and
L.~P.~L.~Colombo\inst{\milanoA, \milanoC}
\and
K.~J.~Andersen\inst{\oslo}
\and
\textcolor{black}{R.~Aurlien}\inst{\oslo}
\and
\textcolor{black}{R.~Banerji}\inst{\oslo}
\and
\textcolor{black}{A.~Basyrov}\inst{\oslo}
\and
M.~Bersanelli\inst{\milanoA, \milanoC, \milanoB}
\and
S.~Bertocco\inst{\triesteB}
\and
M.~Brilenkov\inst{\oslo}
\and
M.~Carbone\inst{\planetek}
\and
H.~K.~Eriksen\inst{\oslo}
\and
J.~R.~Eskilt\inst{\oslo}
\and
\textcolor{black}{M.~K.~Foss}\inst{\oslo}
\and
C.~Franceschet\inst{\milanoA,\milanoC}
\and
\textcolor{black}{U.~Fuskeland}\inst{\oslo}
\and
S.~Galeotta\inst{\triesteB}
\and
M.~Galloway\inst{\oslo}
\and
S.~Gerakakis\inst{\planetek}
\and
E.~Gjerl{\o}w\inst{\oslo}
\and
\textcolor{black}{B.~Hensley}\inst{\princeton}
\and
\textcolor{black}{D.~Herman}\inst{\oslo}
\and
M.~Iacobellis\inst{\planetek}
\and
M.~Ieronymaki\inst{\planetek}
\and
\textcolor{black}{H.~T.~Ihle}\inst{\oslo}
\and
J.~B.~Jewell\inst{\jpl}
\and
\textcolor{black}{A.~Karakci}\inst{\oslo}
\and
E.~Keih\"{a}nen\inst{\helsinkiA, \helsinkiB}
\and
R.~Keskitalo\inst{\nersc}
\and
G.~Maggio\inst{\triesteB}
\and
D.~Maino\inst{\milanoA, \milanoC, \milanoB}
\and
M.~Maris\inst{\triesteB}
\and
B.~Partridge\inst{\haverford}
\and
M.~Reinecke\inst{\mpa}
\and
M.~San\inst{\oslo}
\and
A.-S.~Suur-Uski\inst{\helsinkiA, \helsinkiB}
\and
T.~L.~Svalheim\inst{\oslo}
\and
D.~Tavagnacco\inst{\triesteB, \triesteA}
\and
H.~Thommesen\inst{\oslo}
\and
D.~J.~Watts\inst{\oslo}
\and
I.~K.~Wehus\inst{\oslo}
\and
A.~Zacchei\inst{\triesteB}
}
\institute{\small
Dipartimento di Fisica, Universit\`{a} degli Studi di Milano, Via Celoria, 16, Milano, Italy\goodbreak
\and
INFN, Sezione di Milano, Via Celoria 16, Milano, Italy\goodbreak
\and
Institute of Theoretical Astrophysics, University of Oslo, Blindern, Oslo, Norway\goodbreak
\and
INAF-IASF Milano, Via E. Bassini 15, Milano, Italy\goodbreak
\and
INAF - Osservatorio Astronomico di Trieste, Via G.B. Tiepolo 11, Trieste, Italy\goodbreak
\and
Planetek Hellas, Leoforos Kifisias 44, Marousi 151 25, Greece\goodbreak
\and
Department of Astrophysical Sciences, Princeton University, Princeton, NJ 08544,
U.S.A.\goodbreak
\and
Jet Propulsion Laboratory, California Institute of Technology, 4800 Oak Grove Drive, Pasadena, California, U.S.A.\goodbreak
\and
Department of Physics, Gustaf H\"{a}llstr\"{o}min katu 2, University of Helsinki, Helsinki, Finland\goodbreak
\and
Helsinki Institute of Physics, Gustaf H\"{a}llstr\"{o}min katu 2, University of Helsinki, Helsinki, Finland\goodbreak
\and
Computational Cosmology Center, Lawrence Berkeley National Laboratory, Berkeley, California, U.S.A.\goodbreak
\and
Haverford College Astronomy Department, 370 Lancaster Avenue,
Haverford, Pennsylvania, U.S.A.\goodbreak
\and
Max-Planck-Institut f\"{u}r Astrophysik, Karl-Schwarzschild-Str. 1, 85741 Garching, Germany\goodbreak
\and
Dipartimento di Fisica, Universit\`{a} degli Studi di Trieste, via A. Valerio 2, Trieste, Italy\goodbreak
}

\authorrunning{Paradiso et al.}
\titlerunning{\BP\ cosmological parameters}

\abstract{We present cosmological parameter constraints as estimated
  using the Bayesian \BP\ analysis framework. This method supports
  seamless end-to-end error propagation from raw time-ordered data to
  final cosmological parameters. As a first demonstration of the
  method, we analyze time-ordered \Planck\ LFI observations, combined
  with selected external data (\WMAP\ 33--61\,GHz, \Planck\ HFI DR4 353
  and 857\,GHz, and Haslam 408\,MHz) in the form of pixelized maps which are 
  used to break critical astrophysical degeneracies.  Overall, all results are
  generally in good agreement with previously reported values from
  \Planck\ 2018 and \WMAP, with the largest relative difference for any
  parameter of about $1\,\sigma$ when considering only temperature
  multipoles between $30\le\ell\le600$. In cases where there are
  differences, we note that the \BP\ results are generally slightly
  closer to the high-$\ell$ HFI-dominated \Planck\ 2018 results than
  previous analyses, suggesting slightly less tension between low and
  high multipoles. Using low-$\ell$ polarization information from LFI
  and \WMAP, we find a best-fit value of $\tau=0.066\pm0.013$, which
  is higher than the low value of $\tau=0.051\pm0.006$ derived from
  \Planck\ 2018 and slightly lower than the value of $0.069\pm0.011$
  derived from joint analysis of official LFI and
  \WMAP\ products. Most importantly, however, we find that the
  uncertainty derived in the \BP\ processing is about 30\,\% larger
  than when analyzing the official products, after taking into account
  the different sky coverage. We argue that this is due to
  marginalizing over a more complete model of instrumental and
  astrophysical parameters, and this results in both more reliable and
  more rigorously defined uncertainties. We find that about 2000 Monte
  Carlo samples are required to achieve robust convergence for a
  low-resolution CMB covariance matrix with 225 independent modes, and
  producing these samples takes about eight weeks on a modest
  computing cluster with 256 cores.}


\maketitle

\tableofcontents

\section{Introduction}
\label{sec:introduction}

The cosmic microwave background (CMB) represents one of the most
powerful probes of cosmology available today, as small variations in
the intensity and polarization of this radiation impose strong
constraints on cosmological structure formation processes in the early
universe. The first discovery of these fluctuations was made by
\citet{smoot:1992}, and during the last three decades massive efforts
have been spent on producing detailed maps with steadily increasing
sensitivity and precision \citep[e.g.,][and references
  therein]{bennett2012,debernardis:2000,Louis:2017,Sievers:2013,ogburn:2010,
  planck2016-l01}. These measurements have led to a spectacularly
successful cosmological concordance model called $\Lambda$CDM that
posits that the universe was created during a hot Big Bang about 13.8
billion years ago; that it was seeded by Gaussian random density
fluctuations during a brief period of exponential expansion called
inflation; and that it consists of about 5\,\% baryonic matter, 30\,\%
dark matter, and 65\,\% dark energy. This model is able to describe a
host of cosmological observables with exquisite precision \citep[see
  e.g.][]{planck2016-l06}, even though it leaves much to be desired in
terms of theoretical understanding. Indeed, some of the biggest
questions in modern cosmology revolve around understanding the
physical nature of inflation, dark matter and dark energy, and
billions of dollars and euros are spent on these questions. CMB
observations play a key role in all these studies.

The current state-of-the-art in terms of full-sky CMB observations is
defined by ESA's \Planck\ satellite mission
\citep{planck2013-p01,planck2014-a01,planck2016-l01}, which observed
the microwave sky in nine frequencies, ranging from 30 to 857\,GHz,
between 2009 and 2013. These measurements imposed strong constraints
on the \LCDM\ model, combining information from temperature and
polarization CMB maps with novel gravitational lensing reconstructions
\citep{planck2016-l06}. While the \Planck\ instrument stopped
collecting data already in 2013, the final official \Planck\ data
release took place as recently as 2020 \citep{npipe}, and this
clearly testifies to the significant data analysis challenges
associated with these types of data. Large-scale polarization
reconstruction represents a particularly difficult problem, and
massive amounts of effort have been spent aiming to control all
significant systematic uncertainties
\citep[e.g.,][]{npipe,delouis:2019}.

The next major scientific endeavor for the CMB community is the
search for primordial gravitational waves created during the
inflationary epoch \citep[e.g.,][]{kamionkowski:2016}. Current theories predict that such gravitational
waves should imprint large-scale $B$-mode polarization in the CMB
anisotropies, with a map-domain amplitude no larger than a few tens of
nano Kelvin on degree angular scales. Detecting such a faint signal
requires at least one or two orders of magnitude higher sensitivity
than \Planck, and correspondingly more stringent systematics
suppression and uncertainty assessment.

The main operational goal of the \BP\ project \citep{bp01} is to
translate some of the main lessons learned from \Planck\ in terms of
systematics mitigation into practical computer code that can be used
for next-generation B-mode experiment analysis. And among the most
important lessons learned in this respect from \Planck\ regards the
tight connection between instrument characterization and astrophysical
component separation. Because any CMB satellite experiment in practice
must be calibrated with in-flight observations of astrophysical
sources, the calibration is in practice limited by our knowledge by
the astrophysical sources in question---and this must itself be
derived from the same data set. Instrument calibration and component
separation must therefore be performed jointly, and a non-negligible
fraction of the full uncertainty budget arises from degeneracies
between the two.

The \BP\ project addresses this challenge by constructing a complete
end-to-end analysis pipeline for CMB observation into one integrated
framework that does not require intermediate human intervention. This
is the first complete approach to support seamless end-to-end error
propagation for CMB applications, including full marginalization over
both instrumental and astrophysical uncertainties and their
internal degeneracies; see \citet{bp01,bp11} for further discussion.

For pragmatic reasons, the current \BP\ pipeline has so far only been
applied to the \Planck\ LFI observations, which have significantly
lower signal-to-noise ratio than the \Planck\ HFI observations. The
cosmological parameter constraints derived in the following are
therefore not by themselves competitive in terms of absolute
uncertainties as compared with already published
\Planck\ constraints. Rather, the present analysis focuses primarily
on general algorithmic aspects, and represents a first real-world
demonstration of the end-to-end Bayesian framework that will serve as
a platform for further development and data integration of different
experiments \citep{bp05}.

Noting the sensitivity of large-scale polarization reconstruction 
to systematic uncertainties, we adopt the reionization optical depth $\tau$ as a
particularly important probe of stability and performance of the
\BP\ framework, and aim to estimate $P(\tau\mid\mathbf{d})$ from
\Planck\ \lfi\ and \WMAP\ observations. We also constrain a basic
$6$-parameter \LCDM\ model, combining the \BP\ low-$\ell$ likelihood
with a high-$\ell$ Blackwell-Rao CMB temperature likelihood that for
the first time covers the two first accoustic peaks, or
$\ell\le600$. We eventually also complement this with the
\Planck\ high-$\ell$ likelihood to extend the multipole range to the
full \Planck\ resolution, as well as with selected external non-CMB data
sets.

The structure of the rest of the paper is as follows: In
Sect.~\ref{sec:bp} we review the global \BP\ data model, posterior
distribution and the CMB likelihood, focusing in particular on how
cosmological parameters are constrained in this framework. In
Sect.~\ref{sec:LCDM_constraints} we present \LCDM\ parameter
constraints from \BP\ alone and combined with the \Planck\ high-$\ell$
likelihood. In Sect.~\ref{sec:low_ell_results} we assess the impact
of systematic uncertainties, adopting $\tau$ as a reference
parameter. In Sect.~\ref{sec:convergence}, we provide an assessment of
the Monte Carlo convergence of CMB samples. Finally, we summarize our
main conclusions in Sect.~\ref{sec:conclusions}.

\begin{table}[ht]
  \begingroup
  \newdimen\tblskip \tblskip=5pt
  \caption{Overview of cosmological parameters considered in this
    analysis in terms of mathematical symbol, prior range, and short
    description (see text for details). The top block lists
    the base parameters with uniform priors that are directly sampled in the
    MCMC chains. The lower block contains the main derived parameters.
  }
  \label{tab:params}
   \nointerlineskip
  \vskip -3mm
  \footnotesize
  \setbox\tablebox=\vbox{
    \newdimen\digitwidth
    \setbox0=\hbox{\rm 0}
    \digitwidth=\wd0
    \catcode`*=\active
    \def*{\kern\digitwidth}
    \newdimen\signwidth
    \setbox0=\hbox{-}
    \signwidth=\wd0
    \catcode`!=\active
    \def!{\kern\signwidth}
 \halign{
   \hbox to 0.0cm{#\leaderfil}\tabskip -2em&
      \hfil#\hfil \tabskip 0.5em&
      #\hfil \tabskip 0mm\cr
    \noalign{\doubleline}
      \omit\textsc{Parameter}\hfil&
      \hfil\textsc{Uniform prior}\hfil&
      \hfil\textsc{Definition}\hfil\cr
      \noalign{\vskip 4pt\hrule\vskip 4pt}
      \omit\parbox{3cm}{\textit{Base params}}\cr
      \noalign{\vskip 2pt}
      \hspace{3mm} $\omega_{\mathrm b}\equiv\Omega_{\mathrm b}h^2$ 
      & $\lbrack 0.0005,0.1 \rbrack$& 
      \footnotesize{Baryon density today}\cr
      \hspace{3mm} $\omega_{\mathrm c}\equiv\Omega_{\mathrm c} h^2$ 
      & $\lbrack 0.0001,0.99 \rbrack$& 
      \footnotesize{Cold dark matter density today}\cr
      \hspace{3mm} $100\theta_{\mathrm{MC}}$ & $\lbrack 0.5,10.0 \rbrack$& 
      \footnotesize{$100\,\times$ approximation to $r_{\star}/D_{\mathrm A}$}\cr
      \hspace{3mm} $\tau$ & $\lbrack 0.01,0.8 \rbrack$& 
      \footnotesize{Optical depth of reionization}\cr
      \hspace{3mm} $n_{\mathrm s}$ & $\lbrack 0.9,1.1 \rbrack$& 
      \footnotesize{Scalar index 
      ($k_0=\SI{0.05}{\per\mega\parsec}$)}\cr
      \hspace{3mm} $\ln(10^{10}A_{\mathrm s})$ & $\lbrack 2.7,4.0 \rbrack$& 
      \footnotesize{Log  
      ($k_0=\SI{0.05}{\per\mega\parsec}$)}\cr
      \noalign{\vskip 4pt\hrule\vskip 4pt}
      \omit\parbox{3cm}{\textit{Extensions}}\cr
      \noalign{\vskip 2pt}
      \hspace{3mm} $r$ & $\lbrack 0,3 \rbrack$& 
      \footnotesize{Tensor-to-scalar ratio}\cr
      \noalign{\vskip 4pt\hrule\vskip 4pt}
      \omit\parbox{3cm}{\textit{Derived params}}\cr
      \noalign{\vskip 2pt}
      \hspace{3mm} $\Omega_\Lambda$ & & 
      \footnotesize{Dark energy density}\cr
      \hspace{3mm} $\t_0$ & & 
      \footnotesize{Age of the Universe today (in $\si{\giga\year}$)}\cr
      \hspace{3mm} $\Omega_{\mathrm m}$ & & 
      \footnotesize{Matter density}\cr
      \hspace{3mm} $\sigma_8$ & & 
      \footnotesize{RMS matter fluctuation today}\cr
      \hspace{3mm} $z_{\mathrm{re}}$ & & 
      \footnotesize{Redshift of  half re-ionized}\cr
      \hspace{3mm} $H_0$ & $\lbrack 20,100 \rbrack$ & 
      \footnotesize{Expansion rate in 
      $\si{\kilo\metre\per\second\per\mega\parsec}$}\cr
      \hspace{3mm} $10^9A_{\mathrm s}$ & & 
      \footnotesize{$10^9\,\times$ power at $k_0=\SI{0.05}{\per\mega\parsec}$}\cr
       \hspace{3mm} $10^9A_{\mathrm s}e^{-2\tau}$ & & 
      \footnotesize{Scalar power amplitude}\cr
      \noalign{\vskip 4pt\hrule\vskip 5pt} } }
  \endPlancktable \endgroup
\end{table}

\section{Cosmological parameters and \BP}
\label{sec:bp}

We start by introducing the global \BP\ data model in order to
show how it couples to cosmological parameters through the Gibbs loop;
for a detailed discussion, we refer the interested reader to
\citet{bp01} and references therein. Explicitly, the \BP\ time-ordered
data model reads
\begin{equation}
  \label{eq:data_model}
  \begin{split}
  d_{j,t} &= g_{j,t}{\P}_{tp,j}\left[ {\B}_{pp',j}^{\mathrm{symm}} \sum_c 
  {\M}_{jc}\left(\beta_{p'},\Delta_{bp}^j\right) a_{p'}^c +{\B}_{pp',j}^{\mathrm{asymm}}
  \left(s_{j,t}^{\mathrm{orb}}+s_{j,y}^{\mathrm{fsl}}\right)\right] + \\
  & \quad + s_{j,t}^{\mathrm{1Hz}} + n_{j,t}^{\mathrm{corr}}+n_{j,t}^{\mathrm{w}},
  \end{split}
\end{equation}
where $j$ indicates radiometer; $t$ and $p$ denotes time sample and
pixel on the sky, respectively; and $c$ refers to a given
astrophysical signal component. Further, $d_{j,t}$ denotes the
measured data value in units of $\si{\volt}$; $g_{j,t}$ denotes the
instrumental gain in units of $\si{\volt\per\Kcmb}$; $\P_{tp,j}$ is
the $N_{\mathrm{TOD}}\times 3N_{\mathrm{pix}}$ pointing matrix, where
$\psi$ is the polarization angle of the respective detector with
respect to the local meridian; $\B_{pp',j}$ denotes beam convolution;
$\M_{jc}\left(\beta_{p'},\Delta_{\mathrm{bp}}^j\right)$ denotes
element $(j,c)$ of an $N_{\mathrm{det}}\times N_{\mathrm{comp}}$
mixing matrix, describing the amplitude of the component $c$, as seen
by radiometer $j$ relative to some reference frequency $j_0$; $a_c^p$
is the amplitude of component $c$ in pixel $p$, measured at the same
reference frequency as the mixing matrix $\M$, and expressed in
brightness temperature units; $s_{j,t}^{\mathrm{orb}}$ is the orbital
CMB dipole signal in units of $\si{\Kcmb}$, including relativistic
quadrupole corrections; $s_{j,t}^{\mathrm{fsl}}$ denotes the
contribution from far sidelobes, also in units of $\si{\Kcmb}$;
$s_{j,t}^{\mathrm{1Hz}}$ accounts for elecronic interference with a
1\,Hz period; $n_{j,t}^{\mathrm{corr}}$ denotes correlated
instrumental noise; and $n_{j,t}^{\mathrm{w}}$ is uncorrelated (white)
noise. The free parameters in this equation are
$\lbrace\g,\Delta_{\mathrm{bp}},\n_{\mathrm{corr}},\a_c,\beta\rbrace$. All
the other quantities are either provided as intrinsic parts of the
original data sets, or given as a deterministic function of already
available parameters.

In addition to the parameters explicitly defined by
Eq.~\eqref{eq:data_model}, we include a set of hyper-parameters for each
free stochastic random field in the model. For instance, for the CMB
component map, $\a^{\mathrm{CMB}}$, we define a covariance
matrix $\S$, which is taken to be isotropic. Expanding
$a_p=\sum_{\ell m} a_{\ell m}Y_\ell(p)$ into spherical harmonics,
its covariance matrix may be written as
\begin{equation}
  \label{eq:cov_harm}
  \S_{\ell m,\ell'm'}\equiv\langle a_{\ell m}a^{*}_{\ell'm'}\rangle=C_{\ell}
  \delta_{mm'}\delta_{\ell\ell'},
\end{equation}
where $C_\ell$ is called the angular power spectrum. This function
is itself a stochastic field to be included in the model and fitted to
the data, and, indeed, the angular CMB power spectrum is one of the
most important scientific targets in the entire analysis. We note that
this spectral-domain covariance matrix approach does not apply solely to
astrophysical components, but also to instrumental stochastic fields,
such as correlated noise \citep{bp06} and time-dependent gain fluctuations \citep{bp07}.

In many cases, the power spectrum may be further modelled in terms of
a smaller set of free parameters, $\xi$, defined through some
deterministic function, $C_{\ell}(\xi)$. For the CMB case, $\xi$ is
nothing but the set of cosmological parameters, and the function $C_{\ell}(\xi)$ is
evaluated using a standard cosmological Boltzmann solver, for instance as implemented
in the \texttt{CAMB} code \citep{Lewis:1999bs}. If we now define the
full set of free parameters in the data model as $\omega \equiv
\lbrace\g,\Delta_{\mathrm{bp}},\n_{\mathrm{corr}},a_c,\beta,
C_{\ell}(\xi)\rbrace$, the goal of the current paper is to derive an
estimate of the cosmological parameter posterior distribution
$P(\xi\mid\d)$, marginalized over all relevant astrophysical and
instrumental parameters. In practice, this marginalization is
performed by first mapping the full joint posterior, $P(\omega
\mid\mathbf{d})$, as a function of $C_{\ell}$ through Monte Carlo
sampling, then deriving a $C_{\ell}$-based CMB power spectrum likelihood from
the resulting power spectrum samples, and finally mapping out this
likelihood with respect to cosmological parameters using the
well-established CosmoMC \citep{cosmomc} code. We describe in
Table~\ref{tab:params} the cosmological parameters included in our analysis. 
The rest of this section details the steps involved in establishing 
the CMB likelihood for this step.

\subsection{The \BP\ posterior distribution and Gibbs sampler}

In order to sample from the full global posterior, $P(\omega
\mid\mathbf{d})$, we start with Bayes' theorem,
\begin{equation}
  \label{eq:bayes}
  P(\omega\mid\d)=\frac{P(\d\mid\omega)P(\omega)}{P(\d)}
  \propto \mathcal{L}(\omega)P(\omega),
\end{equation}
where $P(\d\mid\omega)\equiv\mathcal{L}(\omega)$ is called the
likelihood, $P(\omega)$ is called the prior and $P(\d)$ is a
normalization factor usually referred as the ``evidence''. Since the
latter is independent of $\omega$, we ignore this factor in the
following.

The exact form of the likelihood is defined by the data model in
Eq.~\eqref{eq:data_model}, which is given as a linear sum of various
components, all of which are specified in terms of our free parameters
$\omega$. The only term that is not deterministically defined by
$\omega$ is the white noise, $\n^{\mathrm{w}}$, but this is instead
assumed to be Gaussian distributed with zero mean and covariance
$\N^{\mathrm{w}}$. We can therefore write
$\d=\s^{\mathrm{tot}}(\omega)+\n^\mathrm{w}$, where
$\s^{\mathrm{tot}}(\omega)$ is the sum of all model components in
Eq.~\eqref{eq:data_model}, irrespective of their origin, and therefore
$\d-\s^{\mathrm{tot}}(\omega) \propto N(0,\N^{\mathrm{w}})$, where
$N(\mu,\Sigma)$ denotes a multivariate Gaussian distribution with mean
$\mu$ and covariance $\Sigma$. Thus, the likelihood takes the
following form,
\begin{equation}
  \label{eq:bp_like}
  \mathcal{L}(\omega)\propto P(\n^\mathrm{w}\mid\omega)\propto
  e^{-\frac{1}{2}\left(\d-\s^{\mathrm{tot}}_\omega\right)^t(\N^{\mathrm{w}})^{-1}
  \left(\d-\s^{\mathrm{tot}}_\omega\right)}.
\end{equation}
The priors are less well defined, and the current \BP\ processing uses
a mixture of algorithmic regularization priors (e.g., enforcing
foreground smoothness on small angular scales; \citealp{bp13}),
instrument priors (e.g., Gaussian or log-normal priors on the
correlated noise spectral parameters; \citealp{bp06}), and informative
astrophysical priors (e.g., AME and free-free amplitude priors;
\citealp{bp13,bp11}). A full summary of all active priors is provided in
Sect.~8 of \citet{bp01}.

To map out this billon-parameter sized joint posterior distribution,
we employ Gibbs sampling. That is, rather than drawing samples
directly from the joint posterior distribution, $P(\omega
\mid\mathbf{d})$, we draw samples iteratively from all respective
conditional distributions, partitioned into suitable parameter sets. This sampling scheme may be formally
summarized through the following Gibbs chain,
\begin{alignat}{10}
\g &\,\leftarrow P(\g&\,\mid &\,\d,&\, & &\,\xi_n, &\,\Dbp, &\,\a, &\,\beta, &\,C_{\ell})\\
\n_{\mathrm{corr}} &\,\leftarrow P(\n_{\mathrm{corr}}&\,\mid &\,\d, &\,\g, &\,&\,\xi_n,
&\,\Dbp, &\,\a, &\,\beta, &\,C_{\ell})\\
\xi_n &\,\leftarrow P(\xi_n&\,\mid &\,\d, &\,\g, &\,\n_{\mathrm{corr}}, &\,
&\,\Dbp, &\,\a, &\,\beta, &\,C_{\ell})\\
\Dbp &\,\leftarrow P(\Dbp&\,\mid &\,\d, &\,\g, &\,\n_{\mathrm{corr}}, &\,\xi_n,
&\,&\,\a, &\,\beta, &\,C_{\ell})\\
\beta &\,\leftarrow P(\beta&\,\mid &\,\d, &\,\g, &\,\n_{\mathrm{corr}}, &\,\xi_n,
&\,\Dbp, & &\,&\,C_{\ell})\\\label{eq:cmbamp}
\a &\,\leftarrow P(\a&\,\mid &\,\d, &\,\g, &\,\n_{\mathrm{corr}}, &\,\xi_n,
&\,\Dbp, &\,&\,\beta, &\,C_{\ell})\\\label{eq:cmbcl}
C_{\ell} &\,\leftarrow P(C_{\ell}&\,\mid &\,\d, &\,\g, &\,\n_{\mathrm{corr}}, &\,\xi_n,
&\,\Dbp, &\,\a, &\,\beta&\,\phantom{C_{\ell}})&,
\end{alignat}
where $\leftarrow$ indicates drawing a sample from the distribution on
the right-hand side.

Since the main topic of this paper is cosmological parameter
estimation, we summarize here only the CMB amplitude and power
spectrum sampling steps, as defined by Eqs.~\eqref{eq:cmbamp} and
\eqref{eq:cmbcl}, and refer the interested reader to \citet{bp01} and
references therein for discussions regarding the other steps.

As shown by \citet{jewell2004,wandelt2004}, the amplitude distribution
$P(\a\mid\d,\omega\setminus\a)$, i.e. the probability of $\a$ given the data $\d$ 
and the all the model parameters except $\a$, is a multivariate Gaussian with a
mean given by the so-called Wiener filter and an inverse covariance
matrix given by $\S(C_{\ell})^{-1}+\N^{-1}$, where $\S(C_{\ell})$ and $\N$ now are the
total effective signal and noise covariance matrices, respectively. Samples from this distribution may be
drawn by solving the following system of linear equations, typically
using the Conjugate Gradient method \citep{shewchuk:1994},
\begin{equation}
  \label{eq:wiener}
  \begin{split}
    \biggl(\S^{-1} + \sum_{\nu}\M^t_{\nu}\B^t_{\nu}&\N_{\nu}^{-1}\B_{\nu}\M_{\nu}\biggr)\,\a = \\
    \sum_\nu\M_\nu^t&\B_{\nu}^t\N_\nu^{-1}\m_{\nu} 
    + \sum_{\nu}\M_{\nu}^t\B_{\nu}^t\N_{\nu}^{-1/2}\eta_{\nu} +
    \S^{-1/2}\eta_{0}.
  \end{split}
\end{equation}
In this expression, $\M_{\nu}$ is called the mixing matrix, and
encodes the instrument-convolved spectral energy densities of each
astrophysical foreground component, and the $\eta_i$'s are independent
random Gaussian vectors of $N(0,1)$ variates. For further details on
solving this equation, see \citet{eriksen2008,seljebotn:2019,bp01,bp11}.

Sampling from $P(C_{\ell}\mid\d,\omega\setminus C_{\ell})$ is much
simpler, as this is an inverse gamma distribution with $2\ell+1$
degrees of freedom for CMB temperature measurements
\citep{wandelt2004} and a corresponding Wishart distribution for CMB
polarization \citep{larson:2006}. The standard sampling algorithm for
the former of these is simply to draw $2\ell-1$ random variates from a
standard Gaussian distribution, $\eta_i\sim N(0,1)$, and set $C_{\ell}
= \sigma_{\ell} / \sum \eta_i^2$, where $\sigma_\ell = \sum |a_{\ell
  m}|^2$. The generalization to polarization is straightforward.

The above Gibbs algorithm only represents a formal summary of the
algorithm, and in practice we introduce a few important modifications
for computational and robustness reasons. The first modification
revolves around Galactic plane masking. As shown by \citet{bp11}, the
\BP\ CMB reconstruction is not perfect along the Galactic plane. To
avoid these errors from contaminating the CMB power spectrum and
cosmological parameters, we therefore apply a fairly large confidence
mask for the actual CMB analysis. At the same time, the Galactic plane
does contain invaluable information regarding important global
instrumental parameters, for instance the detector bandpasses
\citep{bp09}, and excluding these data entirely from the analysis
would greatly increase the uncertainties on those parameters. For this
reason, we run the analysis in two main stages; we first run the above
algorithm \emph{without} a Galactic mask and setting
$\S_{\mathrm{CMB}}^{-1} = 0$, primarily to estimate the
instrumental and astrophysical parameters; this configuration
corresponds to estimating the CMB component independently in each
pixel without applying any smoothness prior over the full sky. The
cost of setting the power spectrum prior to zero is slightly larger
pixel uncertainties than in the optimal case, as the CMB field is now
allowed to vary almost independently from pixel to pixel. However,
this also ensures that any potential modelling errors remain local,
and are not spread across the sky.

Then, once this main sampling process is done, we \emph{resample} the
original chains with respect to the CMB component by looping through
each main sample, fixing all instrumental and (most of the)
astrophysical parameters, and sampling the CMB-related parameters
again; see \citet{bp11} for full details. For low-resolution CMB
polarization estimation, for which our likelihood relies on a dense
pixel-pixel covariance matrix, the main goal of this stage is simply
to obtain more samples of the same type as above, to reduce the Monte
Carlo uncertainty in the noise covariance matrix
\citep{sellentin2016}. In this case, we therefore simply draw $n$
additional samples from Eq.~\eqref{eq:cmbamp}, fixing both
instrumental and astrophysical parameters, as well as the CMB $a_{\ell
  m}$'s for $\ell>64$. We thereby effectively map out the local
conditional distribution with respect to white noise for each main sample
on large angular scales. We conservatively draw $n=50$ new samples per
main sample in this step, but after the analysis started, we have
checked that as few as 10~sub-samples achieves an equivalent
effect. On the other hand, since the cost of producing one of these
sub-samples is almost two orders of magnitude smaller than producing a
full sample, this additional cost is negligible.

This approach is not suitable for high-resolution CMB temperature
analysis, since we cannot construct a pixel-pixel covariance matrix
with millions of pixels. In this case, we instead use the Gaussianized
Blackwell-Rao estimator \citep{chu2005,rudjord:2009}, which was also
used for CMB temperature analysis up to $\ell\le30$ by
\Planck\ \citep[e.g.,][]{planck2016-l05}. This estimator relies on
proper $C_{\ell}$ samples, and we therefore resample the main chains
once again, but this time we apply the confidence mask and enable the
$C_{\ell}$ sampling step; once again, all instrumental and (most of)
the astrophysical parameters are fixed at their main chain sample
values. Thus, this step includes solving Eq.~\eqref{eq:wiener} with an
inverse noise covariance matrix that is zero in the masked pixels and
a non-local $\S$ covariance matrix, and this translates into a very
high condition number for the coefficient matrix on the left-hand side
\citep{seljebotn:2019}. In fact, the computational cost of a single
CMB temperature power spectrum sample is comparable to the cost of a
full main sample, and we therefore only produce 4000 of
these. Fortunately, as shown in Sect.~\ref{sec:convergence}, this is
sufficient to achieve good convergence up to $\ell\lesssim
700$. However, it does not allow us to explore the low signal-to-noise
regime above $\ell\gtrapprox 800$. For this reason, we conservatively limit the current
\BP\ temperature analysis to $\ell\le 600$, leaving some buffer, and
combine with \Planck\ 2018 results at higher multipoles when
necessary.

\subsection{The \BP\ CMB likelihood}
\label{subsec:cmb_like_bp}

The \BP\ CMB power spectrum likelihood is based on two
well-established techniques, namely brute-force low-resolution
likelihood evaluation on large angular scales for polarization
\citep[e.g.,][]{page2007,planck2016-l05}, and Blackwell-Rao (BR)
estimation for intermediate angular scales for temperature
\citep{chu2005,rudjord:2009,planck2014-a13}. The main variations are
that we employ the signal-to-noise eigenmode compression technique
described by \citet{tegmark1997,gjerlow2015} for the low-resolution
likelihood (to reduce the dimensionality of the covariance matrix, and
therefore the number of Gibbs samples required for convergence), and
that we now are able to use the BR estimator to $\ell\le 600$, not
only $\ell\le200$, as was done in \Planck\ 2018; the main reason for
this is that in the current scheme the CMB sky map samples are drawn
from foreground-subtracted frequency maps (30, 44, 70\,GHz\ldots),
each with a well-defined white noise term, while in the
\Planck\ analysis they were generated from component-separated CMB
maps (\commander, NILC, SEVEM, and SMICA; \citealp{planck2016-l04})
with smoothed white noise terms. In this section, we briefly review
the mathematical backgrounds for each of these two likelihood
approximations, and refer the interested readers to the already
mentioned papers for further details.

\subsubsection{Low-$\ell$ temperature+polarization likelihood}

Starting with the low-resolution case, the appropriate expression for
a multivariate Gaussian likelihood reads
\begin{equation}
  \label{eq:gauss_like}
  P(C_\ell\mid\hat{\s}_{\mathrm{CMB}}) \propto \frac{\exp{(-\frac{1}{2}\hat{\s}_{\mathrm{CMB}}^t
  \left( \S(C_\ell)+\N \right)^{-1}\hat{\s}_{\mathrm{CMB}}})}{\sqrt{|\S(C_\ell)+\N|}},
\end{equation}
where $\hat{\s}_{\mathrm{CMB}}$ represents a CMB-plus-noise map and
$\N$ is its corresponding effective noise covariance map. This
expression has formed the basis of numerous exact CMB likelihood
codes, going at least as far back as \emph{COBE}-DMR
\citep[e.g.,][]{gorski:1994}.

The key novel aspect of the current analysis is simply how to
establish $\hat{\s}_{\mathrm{CMB}}$ and $\N$: In previous analyses,
$\hat{\s}_{\mathrm{CMB}}$ has typically been estimated by
maximum-likelihood techniques, while $\N$ has been estimated through
analytic evaluations that are only able to take into account a rather
limited set of uncertainties, such as white and correlated noise; a
very simplified template-based foreground model; and simple
instrumental models of modes that have poorly measured gains as a consequence 
of the scanning strategy. In contrast, in the current paper both these
quantities are estimated simply by averaging over all available Gibbs
samples,
\begin{align}
  \hat{\s}_{\mathrm{CMB}} &= \langle\s^i_{\mathrm{CMB}}\rangle \\
  \N &= \left\langle\left( \s^i_{\mathrm{CMB}}-\s_{\mathrm{CMB}} \right)
  \left( \s^i_{\mathrm{CMB}}-\s_{\mathrm{CMB}} \right)^t\right\rangle
  \label{eq:postmean}
\end{align}
where brackets indicate Monte Carlo averages. Thus, in this approach
there is no need to explicitly account for each individual source of
systematic effects in the covariance matrix, but they are all
naturally and seamlessly accounted for through the Markov chain
samples.

The main challenge associated with this approach regards how many
samples are required for $\N$ to actually converge. As discussed by
\citet{sellentin2016}, a general requirement is that
$n_{\mathrm{samp}} \gg n_{\mathrm{mode}}$, where $n_{\mathrm{samp}}$
is the number of Monte Carlo samples and $n_{\mathrm{mode}}$ is the
number of modes in the covariance matrix. To establish a robust
covariance matrix, one may therefore either increase
$n_{\mathrm{samp}}$ (at the cost of increased computational costs) or
decrease $n_{\mathrm{mode}}$ (at the cost of increased final
uncertainties). It is therefore of great interest to compress the
relevant information in $\hat{\s}_{\mathrm{CMB}}$ into a minimal set
of modes that capture as much of the relevant information as
possible. In our case, the main cosmological target for the
low-resolution likelihood is the optical depth of reionization,
$\tau$, and the main impact of this parameter on the $C_{\ell}$ power
spectrum for \Planck\ LFI happens in polarization at very low
multipoles, $\ell \lesssim 6-8$, due to the limited sensitivity of the
instrument \citep{planck2016-l05}.

In practice, we compress the information using the methodology
discussed by \citet{tegmark1997}, which isolates the useful modes
through Karhunen-Lo\`eve (i.e., signal-to-noise eigenmode)
compression. Adopting the notation introduced by \citet{gjerlow2015},
we transform the data into a convenient basis through a linear
operator of the form $\bar{\s}=\P\s^{\mathrm{CMB}}$, where the
projection operator is defined as
\begin{equation}
  \label{eq:s2n_operator}
  \P = \lbrack\P_h\left( \S^{1/2}\N^{-1}\S^{1/2} \right)\P_h^t\rbrack_\epsilon\M.
\end{equation}
Here $\P_h$ is an harmonic space truncation operator that retains only
spherical harmonics up to a truncation multipole $\ell_\mathrm{t}$,
$\M$ is a masking operator, and $\lbrack\A\rbrack_\epsilon$ is the set
of eigenvectors of $\A$ with a fractional eigenvalue larger than a
threshold value $\epsilon$. Thus, $\P$ corresponds to a orthonormal
basis on the masked sky that retains primarily multipoles below
$\ell_\mathrm{t}$,\footnote{Note that these modes do have some sensitivity to higher multipoles due to non-orthogonality of the spherical harmonics on a masked sky; the quoted truncation limit is therefore only approximate.} and with a relative
signal-to-noise ratio higher than $\epsilon$. It is important to note
that this projection operator results in an unbiased likelihood
estimator irrespective of the specific values chosen for
$\ell_\mathrm{t}$ and $\epsilon$, and the only cost of choosing
restrictive values for either is just larger uncertainties in the
final results. This is fully equivalent to masking pixels on the sky;
as long as the mask definition does not exploit information in the CMB
map itself, no choice of mask can bias the final results, but only
modify the final error bars. In this paper, we adopt a multipole
threshold of $ \ell_{\mathrm{max}}=8$ and a signal-to-noise threshold
of $10^{-6}$; we apply the R1.8 analysis mask defined by
\citet{planck2016-l05} (with $f_{\mathrm{sky}}=0.68$); and we use the
best-fit \Planck\ 2018 \LCDM\ spectrum to evaluate $\S$. In total,
this leaves 225 modes in $\P$. Determining how many Monte Carlo
samples are required to robustly map out the likelihood for this
number of modes is one of the key results of
Sect.~\ref{sec:low_ell_results}.

\subsubsection{High-$\ell$ temperature likelihood}

\begin{figure*}[t]
	\center
	\includegraphics[width=\linewidth]{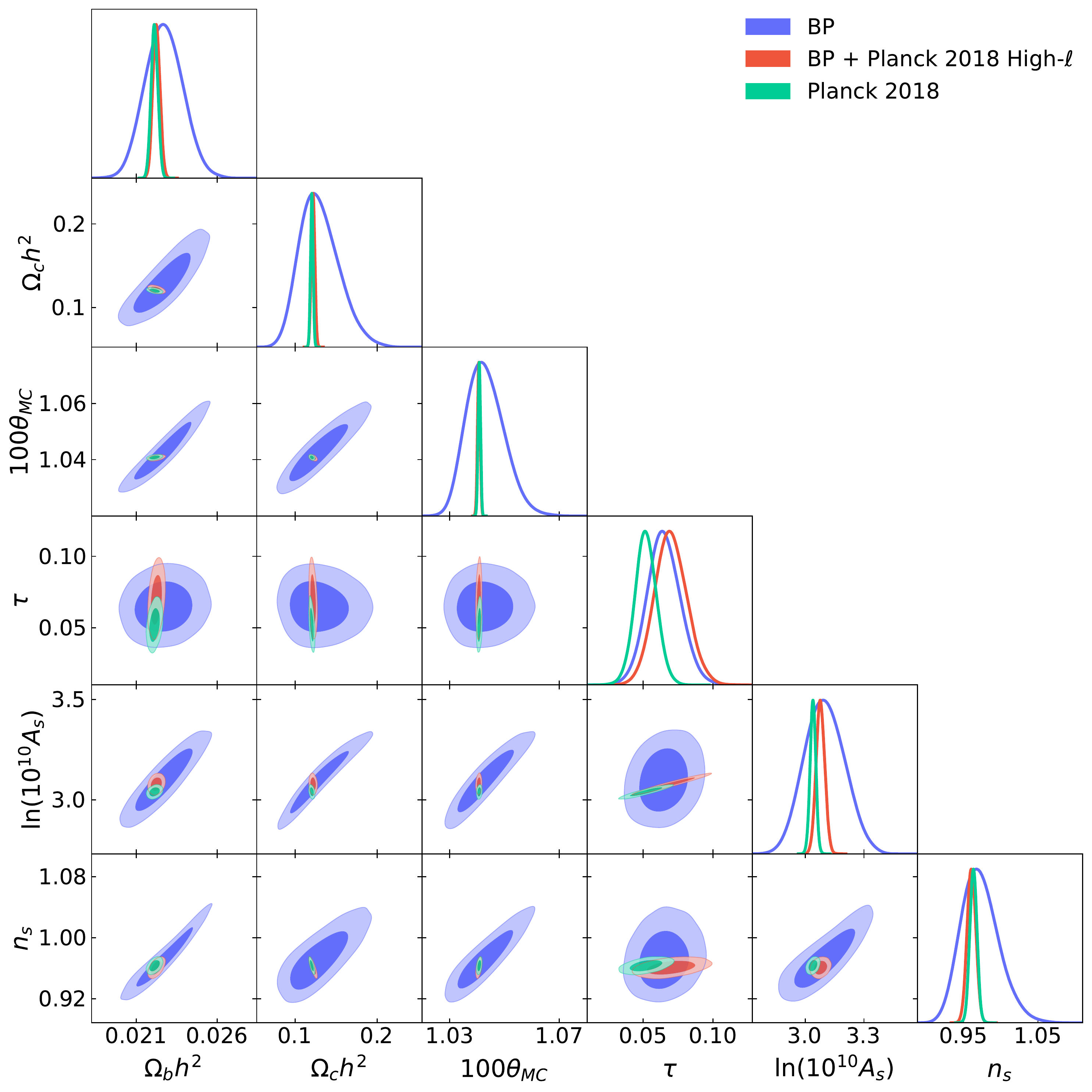}
  \caption{Constraints on the six $\Lambda$CDM parameters from the
    \BP\ likelihood (\emph{blue contours}) using the low-$\ell$ brute-force
    temperature-plus-polarization likelihood for $\ell\le 8$ and the
    high-$\ell$ Blackwell-Rao likelihood for $9\le\ell\le600$. The red
    contours show corresponding constraints when adding the
    high-$\ell$ \Planck\ 2018 $TT$-only likelihood for $601\le\ell\le
    2500$, while green contours show the same for the \Planck\ 2018
    likelihood.}
	\label{fig:full_params}
\end{figure*}

\begin{table}[t]
  \begingroup
  \newdimen\tblskip \tblskip=5pt
  \caption{
    Constraints on the 6 \LCDM\ base parameters with confidence intervals at $68\%$ 
    from CMB data alone and adding lensing + BAO.}
  \label{tab:lcdm+cmb+lens+bao}
   \nointerlineskip
  \vskip -3mm
  \footnotesize
  \setbox\tablebox=\vbox{
    \newdimen\digitwidth
    \setbox0=\hbox{\rm 0}
    \digitwidth=\wd0
    \catcode`*=\active
    \def*{\kern\digitwidth}
    \newdimen\signwidth
    \setbox0=\hbox{-}
    \signwidth=\wd0
    \catcode`!=\active
    \def!{\kern\signwidth}
{\openup 3pt
 \halign{
      #\hfil\tabskip 0.3em&
      \hfil#\hfil\tabskip 0.1em&
      \hfil#\hfil\tabskip 0.1em&
      \hfil#\hfil\cr
    \noalign{\doubleline}
      \omit&
      \omit& 
      \hfil\textsc{\BP\ +}\hfil& 
      \hfil\textsc{\BP\ +}\hfil\cr
      \textsc{Parameter}\hfil & \hfil\textsc{\BP}\hfil & \hfil\textsc{\Planck}\hfil & 
      \hfil\textsc{\Planck\ + Lensing + BAO}\hfil\cr
      \noalign{\vskip 4pt\hrule\vskip 4pt}
       $\Omega_{\mathrm b} h^2   $ & $0.0228^{+0.0011}_{-0.0012}$ & 
      $0.02224\pm 0.00022  $ & $0.02239\pm 0.00020$\cr
       $\Omega_{\mathrm c} h^2   $ & $0.130^{+0.019}_{-0.028}   $  & 
       $0.1218\pm 0.0021 $ & $0.1189\pm 0.0011 $\cr
       $100\theta_{\mathrm{MC}}  $ &  $1.043^{+0.006}_{-0.008}$ & 
       $1.0406\pm 0.0005 $ & $1.0410\pm 0.0004 $\cr
       $\tau$ & $0.065\pm 0.012 $ &  $0.070\pm 0.012  $ & 
      $0.070\pm 0.010 $\cr
       ${\rm{ln}}(10^{10} A_{\mathrm s})$ & $3.10^{+0.10}_{-0.11} $ & 
      $3.078\pm 0.022  $ & $3.071\pm 0.018 $\cr
       $n_{\mathrm s}$ & $0.973^{+0.021}_{-0.029} $ &  $0.961\pm 0.006 $ & 
      $0.967\pm 0.004  $\cr
      \noalign{\vskip 4pt\hrule\vskip 4pt}
      $\Omega_\Lambda$ & $0.63^{+0.14}_{-0.08}$ & $0.673\pm 0.014$ & $0.691\pm 0.007$\cr
      $t_0$ & $13.7^{+0.3}_{-0.2} $  & $13.83\pm 0.04   $ & $13.79\pm 0.03$\cr
      $\Omega_{\mathrm m}$ & $0.37^{+0.08}_{-0.14} $ & $0.327\pm 0.014 $ & $0.309\pm 0.007$\cr
      $\sigma_8$ & $0.87^{+0.12}_{-0.14} $ & $0.830\pm 0.010  $ & $0.819\pm 0.007$\cr
      $z_{\mathrm re}$ & $8.8\pm 1.2  $  & $9.2\pm 1.1 $ & $9.2\pm 0.9 $\cr      
      $H_0$ &  $65.9^{+4.3}_{-5.6} $ & $66.6\pm 0.9  $ & $67.8\pm 0.5  $\cr
      $10^9A_{\mathrm s}e^{-2\tau}$ & $1.96^{+0.17}_{-0.22} $  & $1.888\pm 0.010 $ & $1.875\pm 0.006$\cr
      \noalign{\vskip 4pt\hrule\vskip 5pt} } }}
  \endPlancktablewide \endgroup
\end{table}

For high-$\ell$ temperature analysis, we exploit the Blackwell-Rao
(BR) estimator \citep{chu2005}, which has been demonstrated to work
very well for high signal-to-noise data \citep{eriksen:2004}. This is
the case for the \BP\ temperature power spectrum below $\ell \lesssim
700$, whereas the signal-to-noise ratio for high-$\ell$ polarization is
very low everywhere, even when combining \lfi\ and \WMAP\ data.

In practice, we employ the Gaussianized Blackwell-Rao estimator (GBR),
as presented in \citet{rudjord:2009} and used by
\Planck\ \citep{planck2016-l05}, in order to reduce the number of
samples required to achieve good convergence at high
multipoles. In this approach, the classical Blackwell-Rao estimator is
first used to estimate the univariate $C_{\ell}$ likelihood for each
multipole separately,
\begin{equation}
  P(C_{\ell}\mid\s^{\mathrm{CMB}}) =
  \sum_{i = 1}^{n_{\mathrm{samp}}} \frac{\exp({-\frac{2\ell+1}{2}\frac{\sigma^i_{\ell}}{C_{\ell}}})}{|C_{\ell}|^{\frac{2\ell+1}{2}}},
\end{equation}
where $\sigma^i_{\ell} \equiv \sum_{m} |s^i_{\ell m}|^2/(2\ell+1)$ is
the observed power spectrum of the $i$'th Gibbs sample CMB sky map,
$\s^{\mathrm{CMB}}$. This distribution is used to define a
Gaussianizing change-of-variables, $x_{\ell}(C_{\ell})$,
multipole-by-multipole by matching differential quantiles between the
observed likelihood function and a standard Gaussian distribution. The
final likelihood expression may then be evaluated as follows,
\begin{equation}
  \label{eq:GBR}
  P(C_\ell\mid\d)\approx \left(\prod_\ell\frac{\partial C_\ell}{\partial x_\ell}
  \right)^{-1} e^{-\frac{1}{2}(\x-\mu)^T\Cp^{-1}(\x-\mu)},
\end{equation}
where $\x=\lbrace x_\ell(C_{\ell})\rbrace$ is the vector of transformed
input power spectrum coefficients; $\partial C_\ell/\partial x_\ell$
is the Jacobian of the transformation; and the mean
$\mu=\lbrace\mu_\ell\rbrace$ and covariance matrix $\Cp_{\ell\ell'}=
\langle (x_\ell-\mu_\ell)(x_{\ell'}-\mu_{\ell'})\rangle$ are estimated
from the Monte Carlo samples after Gaussianization with the same
change-of-variables. This expression is by construction exact for the
full-sky and uniform noise case, due to the diagonal form of the noise
covariance matrix, and consequently the full expression factorizes in
$\ell$. For real-world analyses that include sky cuts, anisotropic noise
and systematic uncertainties it is strictly speaking an approximation,
but as shown by \citet{rudjord:2009}, it is an excellent approximation
even for relatively large sky cuts. Furthermore, any differences
induced by additional instrumental systematic error propagation are
small compared to the effect of the Galactic mask, which totally
dominates the sample variance component of the high-$\ell$ temperature
likelihood. In this paper, we derive $\Lambda$CDM cosmological
parameters using the Gaussianized GBR estimator using the multipole
range $9\le\ell\le 600$. Additional details can be found in
\citet{bp01} and \citet{bp11}.

\begin{figure}[t]
	\center
	\includegraphics[width=\linewidth]{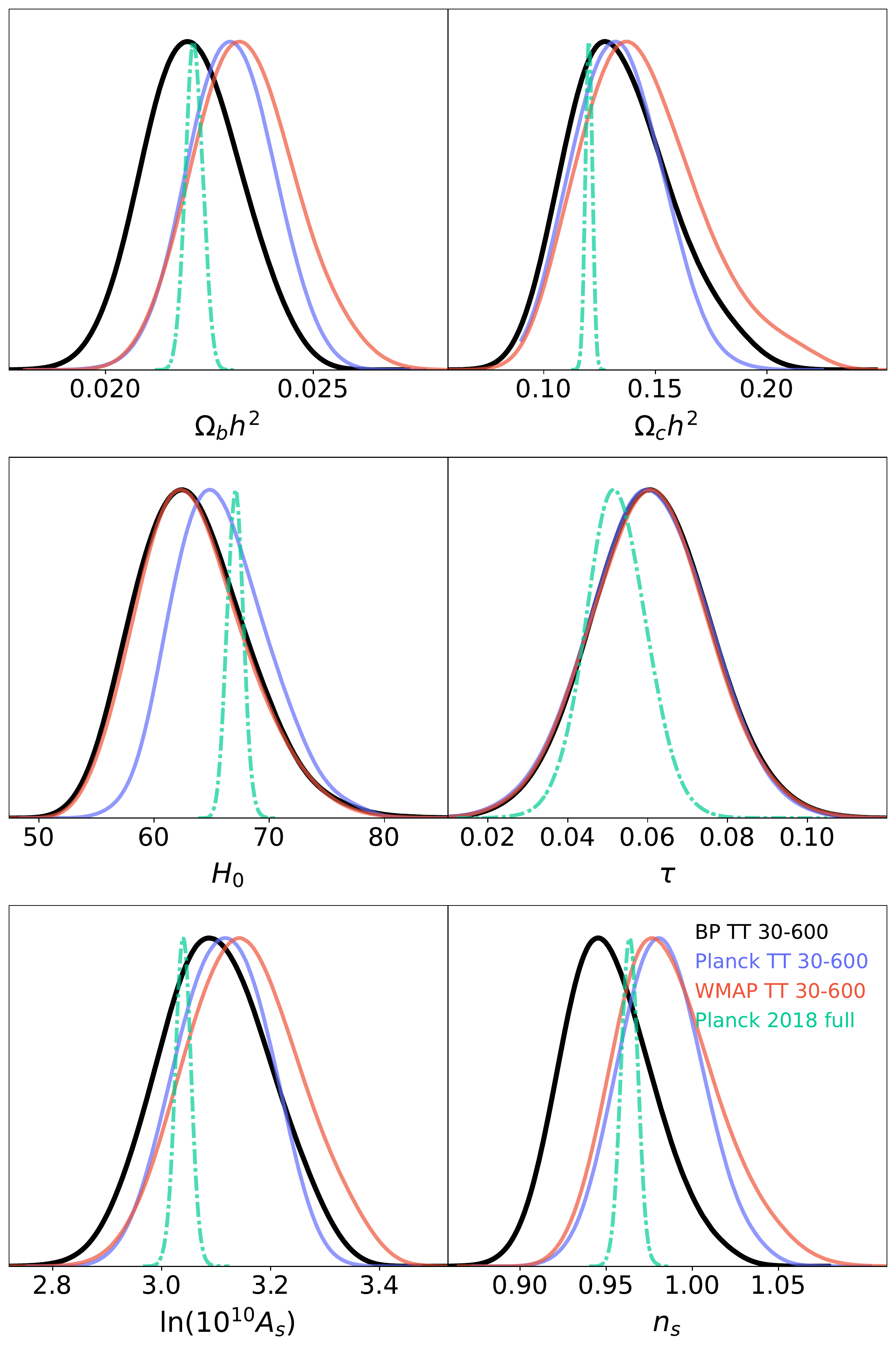}
	\caption{Comparison between \LCDM\ parameters derived using
          $TT$-only between $30\le\ell\le600$ for
          \BP\ (\emph{black}), \Planck\ 2018 (\emph{blue}), and
          \WMAP\ (\emph{red}).  All these cases include a Gaussian
          prior of $\tau=0.06\pm0.015$.  For comparison, the full
          \Planck\ 2018 estimates are shown as dot-dashed green
          distributions.}
	\label{fig:GBR-600}	
\end{figure}

\subsection{CAMB and CosmoMC}
Final cosmological parameters are sampled with \texttt{CosmoMC}
\citep{cosmomc}, using the above likelihoods as inputs. This code
implements a Metropolis-Hastings algorithm to efficiently probe the
whole parameter space, using various speed-up and tuning methods
\citep{neal2005,lewis2013b}. In our analysis, we run eight chains
until they reach convergence, as defined by a Gelman-Rubin statistic
of $R-1<0.01$ \citep{gelman:1992}, while discarding the first 30\,\%
of each chain as burn-in. This is due to the way \texttt{CosmoMC}
learns an accurate orthogonalization and proposal distribution for the
parameters from the sample covariance of the previous samples. In
general, quoted error bars correspond to 68\,\% confidence ranges,
except in cases for which a given parameter is consistent with a hard
prior boundary (such as the tensor-to-scalar ratio, $r$), in which
case we report upper 95\,\% confidence limits. 

\section{Six-parameter \LCDM\ constraints}
\label{sec:LCDM_constraints}

We are now ready to present standard $\Lambda$CDM cosmological
parameter constraints as derived from the \BP\ likelihood, and we
compare these with previous estimates from \Planck 2018
\citep{planck2016-l05}. The main results are shown in
Fig.~\ref{fig:full_params} in terms of one- and two-dimensional
marginal posterior distributions of the six \LCDM\ base parameters for
three different cases. The blue contours show results derived from
\BP\ alone, using only temperature information up to $\ell\le600$ and
polarization information between $2\le\ell\le 8$, while red contours
show corresponding results when the temperature multipole range is
extended with the \Planck\ 2018 $TT$ likelihood\footnote{We adopt the
  public \Planck\ 2018 likelihood code (PLC; version 3.0) when
  extending the \BP\ likelihood and including lensing and BAO
  constraints.} between $601\le\ell\le2500$. Finally, the green contours
show the full \Planck\ 2018 posterior distributions. The \BP\ results
are summarized in terms of posterior means and standard deviations in
Table~\ref{tab:lcdm+cmb+lens+bao}, where we also report constraints
when including CMB lensing and baryonic
acoustic oscillations (BAO); see \citep{planck2013-p11,
  planck2014-a15} for corresponding \Planck\ analyses.

\begin{figure}[ht]
	\center
	\includegraphics[width=\linewidth]{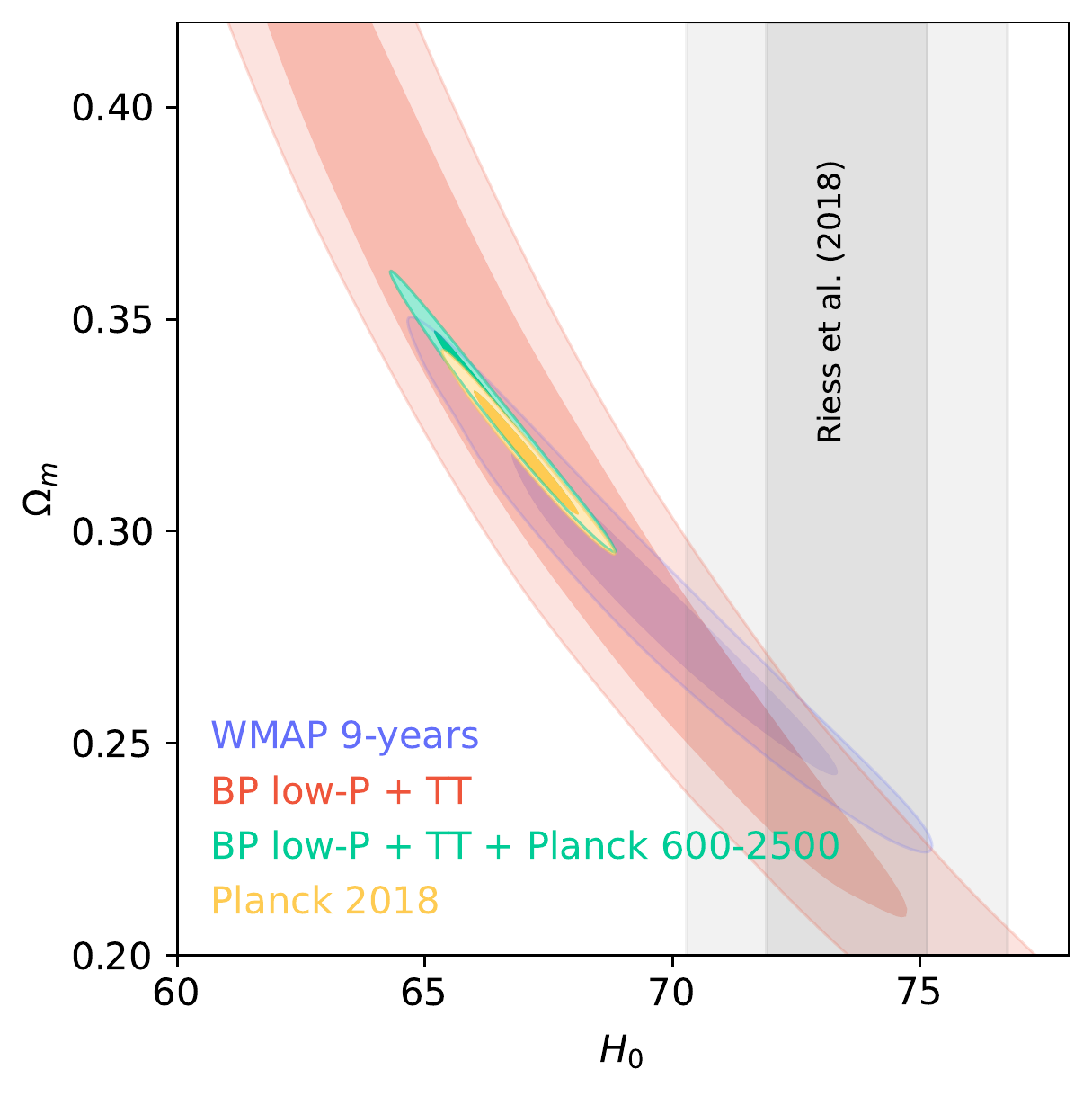} \\
	\includegraphics[width=\linewidth]{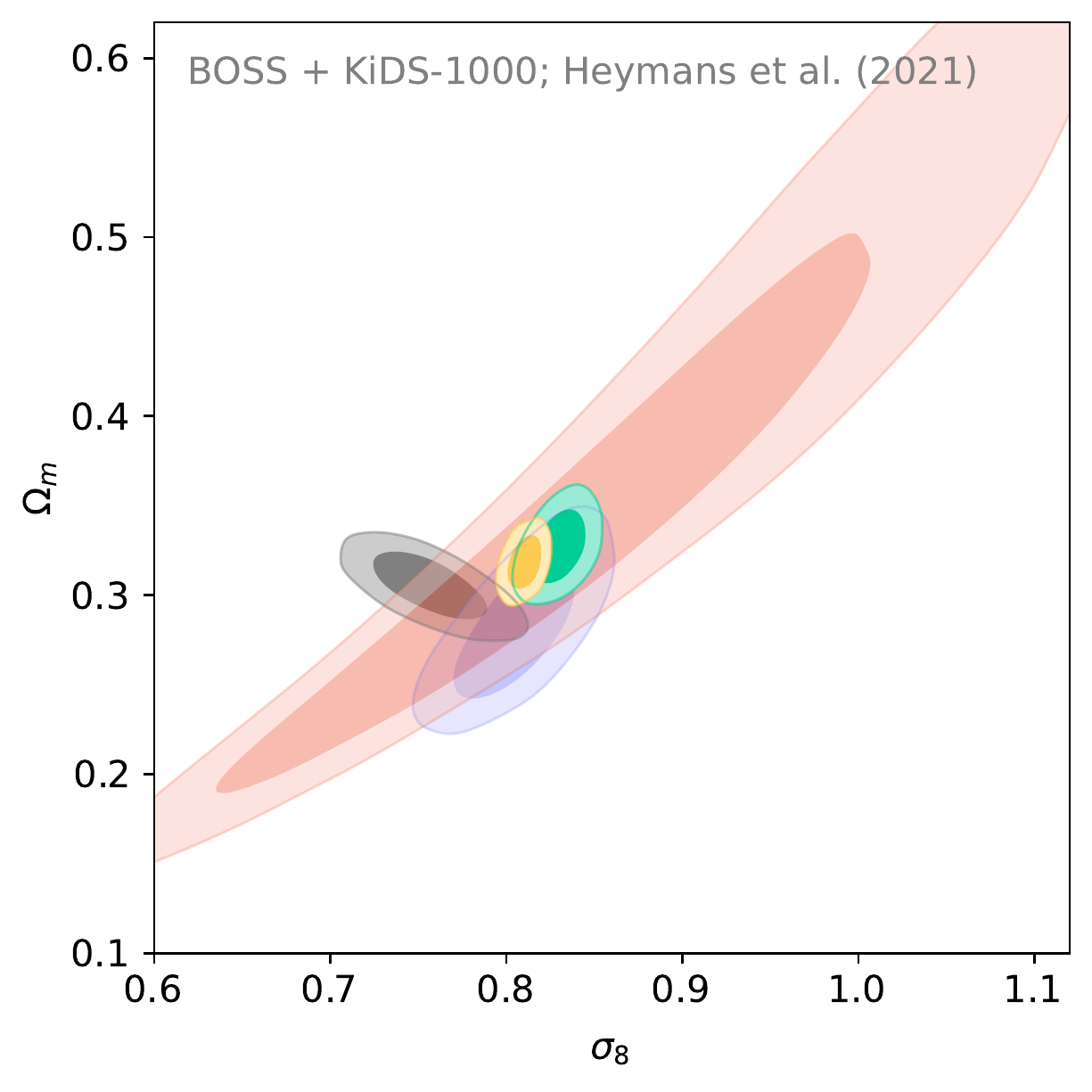}
	\caption{Two-dimensional marginal posterior distributions for the parameter pairs 
	$H_0$--$\Omega_{\mathrm m}$ (\emph{top}), and $\sigma_8$--$\Omega_{\mathrm m}$ (\emph{bottom}) 
	as computed with the \BP-only likelihood (\emph{red}); the
        \BP\ likelihood extended with the \Planck\ 2018 high-$\ell$
        $TT$ likelihood (\emph{green}); the full \Planck\ 2018 likelihood 
	(\emph{yellow}); the \WMAP\ likelihood (\emph{blue}); and, for the 
	\emph{bottom} figure, the joint cosmic shear and galaxy clustering
	likelihood from KiDS-1000 and BOSS \citep[][\emph{gray}]{KiDS2021}.}
	\label{fig:s8_omb}
\end{figure}

Overall, we observe excellent agreement between the various cases, and
the most discrepant parameter is the optical depth of reionization,
for which the \BP\ result ($\tau=0.065\pm0.012$) is higher than the
\Planck\ 2018 constraint ($\tau=0.051\pm0.06$) by roughly
$1\,\sigma$. In turn, this also translates into a higher initial
amplitude of scalar perturbations, $A_{\mathrm{s}}$, by about
$1.5\,\sigma$. At the same time, it is important to note that the
high-$\ell$ information from the HFI-dominated \Planck\ 2018
likelihood plays a key role in constraining all parameters (except
$\tau$), by reducing the width of each marginal distribution by a
factor of typically 5--10. As such, the good agreement seen in
Fig.~\ref{fig:full_params} is not surprising, but rather expected from
the high level of correlations between the input datasets.

It is therefore interesting to assess agreement between the various
likelihood using directly comparable datasets, and such a comparison
is shown in Fig.~\ref{fig:GBR-600}. In this case, we show constraints
derived using only $TT$ information between $30\le\ell\le600$,
combined with a Gaussian prior of $\tau=0.060\pm0.015$. The solid
lines show results for \BP\ (black), \Planck\ 2018 (blue), and
\WMAP\ (red), respectively, while the dashed-dotted green line for
reference shows the same \Planck\ 2018 constraints as in
Fig.~\ref{fig:full_params} derived from the full likelihood.

Taken at face value, the agreement between the three datasets appears
reasonable in this directly comparable regime, as the the most
discrepant parameters are $\Omega_{\mathrm{b}}h^2$ and $H_0$, which
both differs by about $1\,\sigma$ between \BP\ and \Planck\ 2018 and
\WMAP. However, it is important to note that all three of these datasets
are nominally cosmic variance limited in the multipole range between
$30\le\ell\le 600$, and therefore one should in principle expect
perfect agreement between these distributions, and that is obviously
not the case. Some of these discrepancies can be explained in terms of
different masking, noting that the effective sky fraction of the \BP\,
\Planck\ 2018, and \WMAP\ likelihoods are about 63, 65, and 75\,\%,
respectively. However, as shown by \citet{planck2016-l05}, such small
variations are not by themselves large enough to move the main
cosmological parameters by as much as $1\,\sigma$.

It is therefore likely the actual data processing pipelines used to
model and propagate astrophysical and instrumental systematic errors
play a significant role in explaining these differences. In this
respect, we make two interesting observations. First of all, we note
that \BP\ pipeline fundamentally differs from the two previous
pipelines from a statistical point of view, as it is the first
pipeline to implement true end-to-end Bayesian modelling that
propagate all sources of astrophysical and instrumental systematic
uncertainties to the final cosmological parameters; in comparison, the
other two pipelines both rely on a mixture of frequentist and Bayesian
techniques that are only able to propagate a subset of all
uncertainties. Second, we note that the low-$\ell$ LFI-dominated
\BP\ results are for several parameters more consistent with the
high-$\ell$ HFI-dominated \Planck\ 2018 results than the two previous
pipelines; specifically, this is the case for $\Omega_\mathrm{b}h^2$,
$\Omega_\mathrm{c}h^2$, and $A_{\mathrm{s}}$, while for $H_0$, the
\Planck\ 2018 low-$\ell$ likelihood is slightly closer to its
high-$\ell$ results, while \BP\ and \WMAP\ are identical. Finally, for
$n_\mathrm{s}$ all three pipelines result in comparable agreement with
the high-$\ell$ result in terms of absolute discrepancy, but with a
different sign; \BP\ prefers a stronger tilt than either \Planck\ 2018
or \WMAP. All in all, we conclude that there seems to be slightly less
internal tension between low and high multipoles when using the
\BP\ likelihood. Still, the main conclusion from this analysis is that
all these differences are indeed small in an absolute sense, and
subtle differences at the $1\,\sigma$ level for $30\le\ell\le 600$ do
not represent a major challenge for the overall cosmological
parameters derived from the full \Planck\ 2018 data, as explicitly
shown in Fig.~\ref{fig:full_params}.

\begin{figure*}[t]
  \center
  \includegraphics[width=0.49\linewidth]{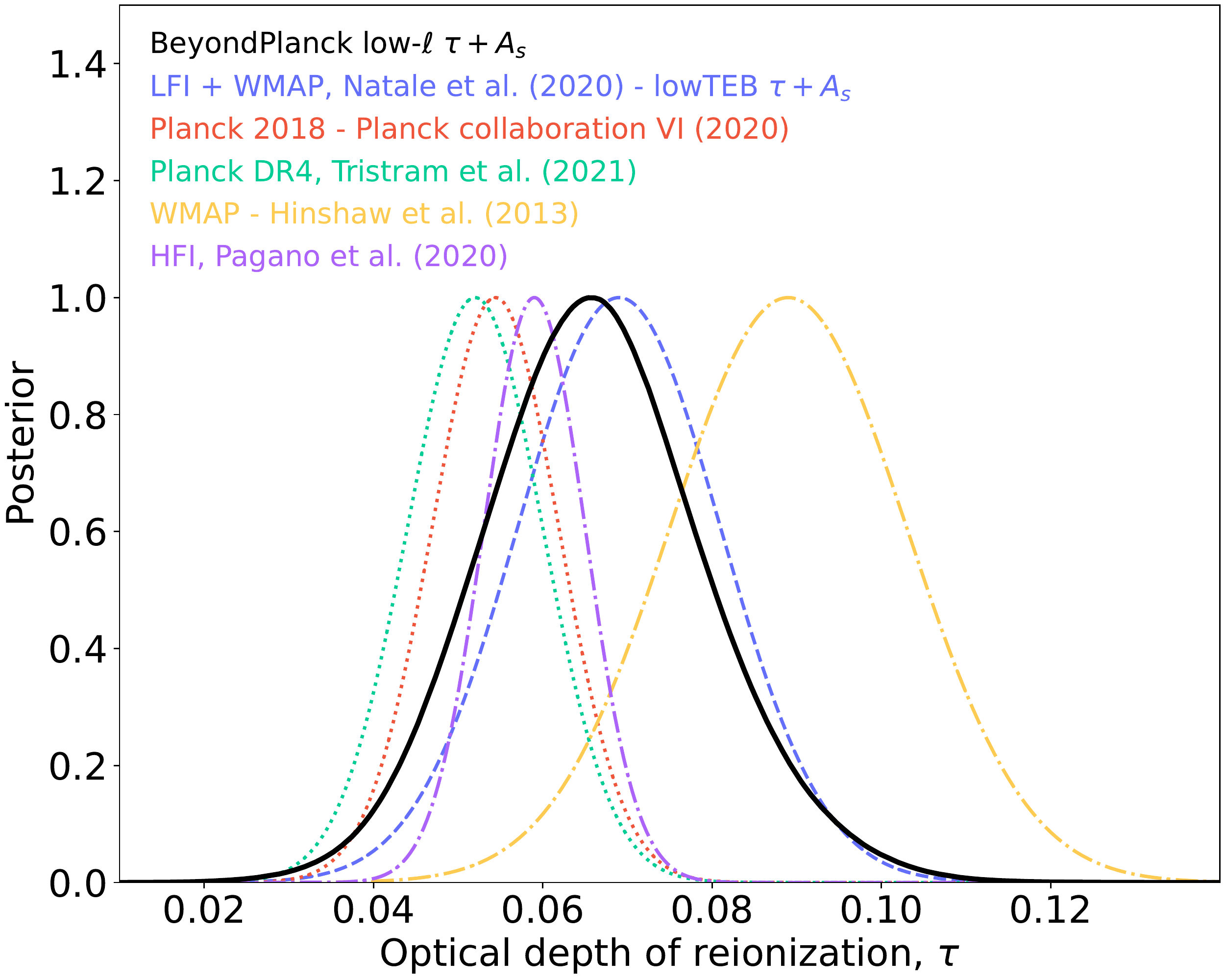} 
  \includegraphics[width=0.49\linewidth]{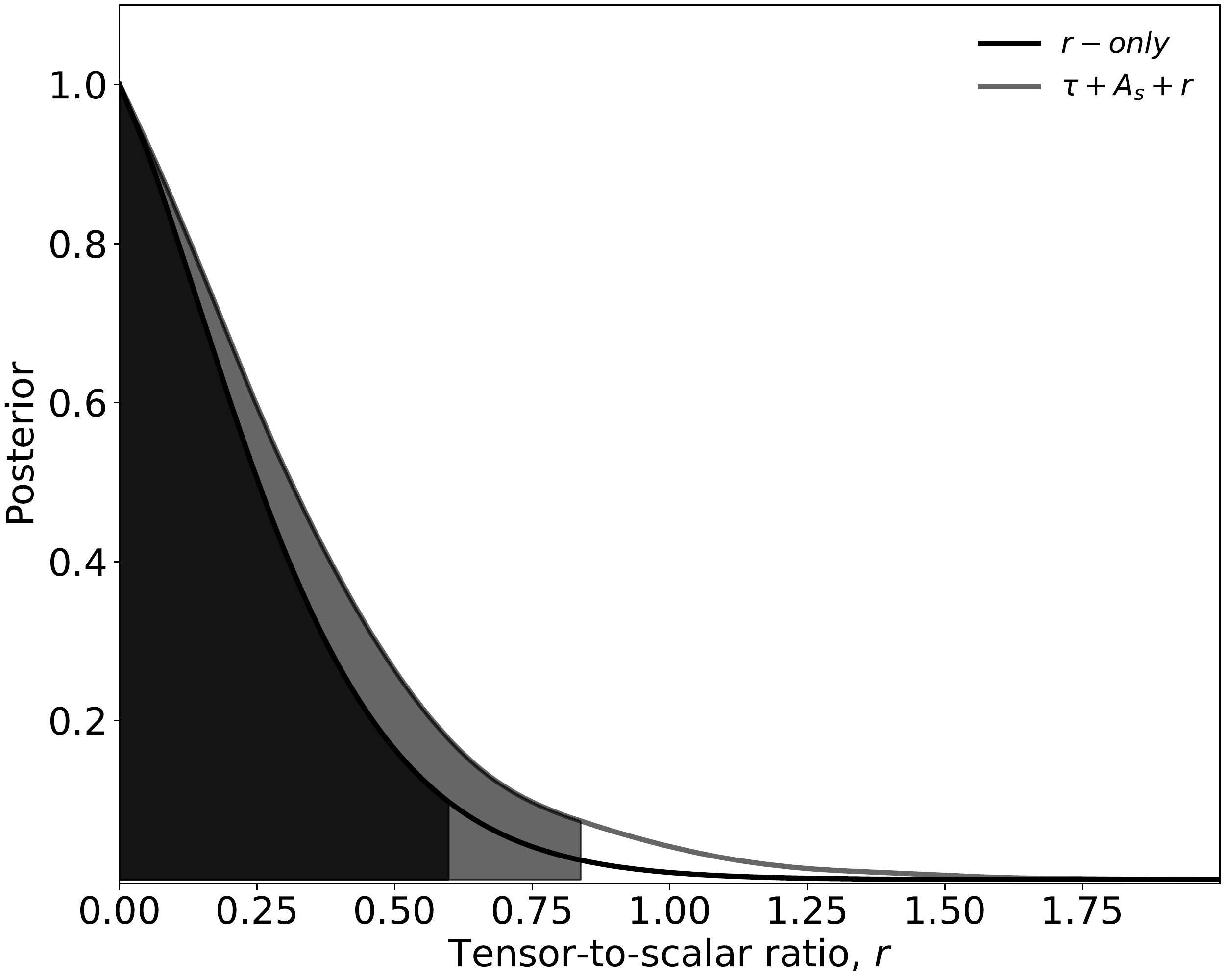}  
  \caption{\emph{Left panel}: Comparison of marginal posterior
    distributions of the reionization optical depth from \Planck\ 2018
    (\emph{red, dotted}; \citealp{planck2016-l06}), 9-year
    \WMAP\ (\emph{yellow, dot-dashed}; \citealp{hinshaw2012}),
    \Planck\ DR4 (\emph{cyan, dotted}; \citealp{tristram:2021}),
    \Planck\ \hfi\ (\emph{purple, dot-dashed}; \citealp{pagano:2020}),
    \WMAP\ \emph{Ka}--\emph V and LFI 70\,GHz (fitting $\tau +
    A_{\mathrm s}$; \citealp{natale:2020}; \emph{blue, dashed}); and
    \BP\ using multipoles $\ell=2$--8, marginalized over the scalar
    amplitude $A_{\mathrm s}$ (\emph{black}).  \emph{Right panel}:
    Corresponding marginal \BP\ tensor-to-scalar ratio posteriors
    derived using $\mathrm{BB}$ multipoles between $\ell=2$--8,
    marginalized over the scalar amplitude $A_{\mathrm s}$
    (\emph{gray}), and by fixing all the \LCDM\ parameters to their
    best-fit values. (\emph{black}). The filled region corresponds to
    the 95\% confidence interval.}
    \label{fig:tau}
\end{figure*}

\begin{table*}[t]       
  \begingroup                                                                            
  \newdimen\tblskip \tblskip=5pt
  \caption{Summary of cosmological parameters dominated by large-scale
    polarization and goodness-of-fit statistics. Columns list, from
    left to right, 1) analysis name; 2) basic data sets included in
    the analysis; 3) effective accepted sky fraction; 4) posterior
    mean estimate of the optical depth of reionization with 68\,\%
    error bars; 5) upper limit on tensor-to-scalar ratio at 95,\%
    confidence; 6) $\chi^2$ goodness-of-fit statistic as measured in
    terms of probability-to-exceed; and 7) primary reference.
    \label{tab:cospar_pol}}
  \nointerlineskip                                                                                                                                                                                     
  \vskip -2mm
  \footnotesize                                                                                                                                      
  \setbox\tablebox=\vbox{ %
  \newdimen\digitwidth                                                                                                                          
  \setbox0=\hbox{\rm 0}
  \digitwidth=\wd0
  \catcode`*=\active
  \def*{\kern\digitwidth}
  \newdimen\signwidth
  \setbox0=\hbox{+}
  \signwidth=\wd0
  \catcode`!=\active
  \def!{\kern\signwidth}
  \newdimen\decimalwidth
  \setbox0=\hbox{.}
  \decimalwidth=\wd0
  \catcode`@=\active
  \def@{\kern\signwidth}
  \halign{ \hbox to 1.8in{#\leaderfil}\tabskip=1.0em&
    \hfil$#$\tabskip=1.5em&
    \hfil$#$\hfil\tabskip=1em&
    \hfil$#$\hfil\tabskip=1em&
    \hfil$#$\hfil\tabskip=0.5em&
    \hfil$#$\hfil\tabskip=1em&
    $#$\hfil\tabskip=0em\cr
  \noalign{\doubleline}
  \omit{\sc Analysis Name}\hfil& \omit{\sc Data Sets}\hfil&\hfil f^{\mathrm{pol}}_{\mathrm{sky}}\hfil&\hfil \tau\hfil&\hfil r^{BB}_{95\,\%}\hfil&\omit\hfil $\chi^2$ {\sc PTE}\hfil&\omit{\sc Reference}\hfil\cr
  \noalign{\vskip 3pt\hrule\vskip 5pt}
  \WMAP\ 9-yr& \omit \WMAP\ \emph{Ka}--\emph V\hfil& 0.76&  0.089\pm0.014& & & \omit\citet{hinshaw2012}\hfil\cr
    Natale et al.& \omit LFI 70, \WMAP\ \emph{Ka}--\emph V\hfil& 0.54&  0.069\pm0.011& & & \omit\citet{natale:2020}\hfil\cr
  \Planck\ 2018& \omit HFI 100$\times$143\hfil& 0.50&  0.051\pm0.009& <0.41 & & \omit\citet{planck2016-l05}\hfil\cr
  \srollTwo & \omit HFI 100$\times$143\hfil& 0.50&  0.059\pm0.006& & & \omit\citet{pagano:2020}\hfil\cr    
  \npipe\ (\commander\ CMB)& \omit LFI+HFI\hfil& 0.50&  0.058\pm0.006&
  < 0.16 & & \omit\citet{tristram:2020}\hfil\cr
  \noalign{\vskip 3pt}   
  {\bf \BP}, $\ell=2$--8 & \omit \bf LFI, \WMAP\ \emph{Ka}--\emph V\hfil&
  \bf 0.68&  \bf 0.066\pm{0.013}& \bf <0.6& \bf 0.32& \textbf{This paper}\cr 
  \BP, $\ell=3$--8 & \omit LFI, \WMAP\ \emph{Ka}--\emph V\hfil& 0.68&  0.066\pm{0.014}& <0.8& 0.32& \textrm{This paper}\cr 
    \noalign{\vskip 3pt\hrule\vskip 5pt}   
  }}
  \endPlancktablewide                                                                                                                                            
  \endgroup
\end{table*}

Before concluding this section, we comment on two important
cosmological parameters that have been the focus of particularly
intense discussion after the \Planck\ 2018 release, namely the Hubble
expansion parameter, $H_0$, and the RMS amplitude of scalar density
fluctuations, $\sigma_8$. Figure~\ref{fig:s8_omb} shows
two-dimensional marginal distributions for $H_0$--$\Omega_\mathrm{m}$
and $\sigma_8$--$\Omega_\mathrm{m}$, respectively, for various data
combinations. Here we see that \BP\ on its own is not able to shed new
light on the either of the two controversies, due to its limited
angular range. When combining with high-$\ell$ \Planck\ 2018
information, however, we see that \BP\ prefers an even slightly lower
mean value of $H_0$ than \Planck 2018, although also with a slightly larger
uncertainty. The net discrepancy with respect to \citet{Riess2018a} is
therefore effectively unchanged.

The same observation holds for $\sigma_8$, for which \BP\ prefers a
higher mean value than \Planck, increasing the absolute discrepancy
with cosmic shear and galaxy clustering measurements from
\citet{KiDS2021}. In this case, we see that \BP\ prefers an even
higher value than \Planck, by about $1.5\,\sigma$, further increasing
the previously reported tension with late-time measurements. This
difference with respect to \Planck\ is driven by the higher value of
$\tau$, as already noted in Fig.~\ref{fig:full_params}.

\begin{figure}[t]
	\center
	\includegraphics[width=\linewidth]{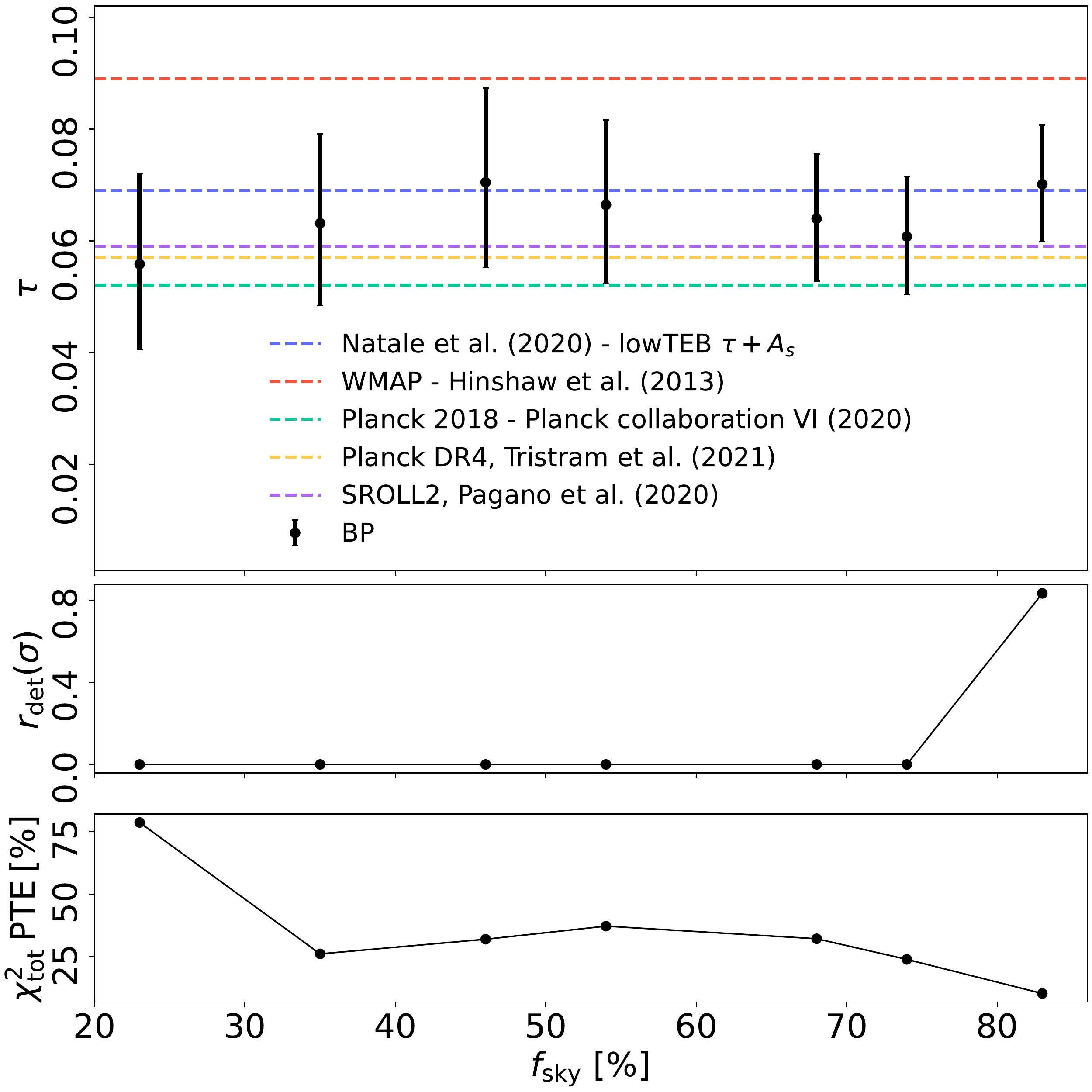}
  	\caption{Low-$\ell$ likelihood stability as a function of sky
          fraction. All results are evaluated adopting the same series
          of \lfi\ processing masks as defined by
          \citep{planck2016-l05}. From top to bottom, the three panels
          show 1) posterior $\tau$ estimate; 2) posterior $r$
          estimate, expressed in terms of a detection level with
          respect to a signal with vanishing B-modes in units of
          $\sigma$; and 3) $\chi^2$ PTE evaluated for the best-fit
          power spectrum in each case.}
	\label{fig:lowl_fsky}
\end{figure}

\section{Large-scale polarization and the optical depth of reionization}
\label{sec:low_ell_results}

As discussed by \citet{bp01}, the main purpose of the \BP\ project was
not to derive new state-of-the-art $\Lambda$CDM parameter constraints,
for which, as we have seen above, \Planck\ HFI data are
essential. Rather, the main motivation behind this work was to
develop a novel and statistically consistent Bayesian end-to-end
analysis framework for past, current and future CMB experiments, with
a particular focus on next-generation polarization experiments. As
such, the single most important scientific target in all of this work
is the optical depth of reionization, $\tau$, which serves as an
overall probe of the efficiency of the entire framework. We are now
finally ready to present the main results regarding this parameter in
this section.

In the left panel of Fig.~\ref{fig:tau} we show the marginal posterior
distribution for $\tau$ as derived from the low-$\ell$ \BP\ likelihood
alone (black curve), and compare this with corresponding previous
estimates from the literature \citep{hinshaw2012, planck2016-l06,
  natale:2020,pagano:2020}. We note, however, that making head-to-head
comparisons between all of these is non-trivial, as the reported
parameters depend on different assumptions and data combinations. For
example, \citet{pagano:2020} considers a likelihood that includes
high-$\ell$ temperature information and marginalizes over all
\LCDM\ parameters, whereas \citet{natale:2020} considers only
low-$\ell$ polarization and marginalizes over only a small set of
strongly correlated parameters, i.e., $A_{\mathrm{s}}$ and/or
$r$. Taking into account the fact that \citet{natale:2020} analyzes
the official LFI and \WMAP\ products jointly, we choose to tune our
analysis configuration to them, to facilitate a head-to-head
comparison for the most relevant case. Corresponding numerical values
are summarized in Table~\ref{tab:cospar_pol}.

We see that the \BP\ polarization-only estimate is in reasonable
agreement with the Natale et al.\ result based on the official LFI and
\WMAP\ products, with an overall shift of about
$0.2\,\sigma$. However, there are two important differences to note in
this regard. First, the \BP\ mean value is slightly lower than the
LFI+\WMAP\ value, and therefore in slightly better agreement with the
HFI-dominated results. Second, and more importantly, we see that the
\BP\ uncertainty is \emph{larger} for \BP\ than LFI+\WMAP\, despite
the fact that its sky fraction is larger (68 versus 54\,\%). Since the
uncertainty on $\tau$ scales roughly inversely proportionally with the
square root of the sky fraction\footnote{This assumption has been verified
by simulating two sets of $1000$ CMB plus noise maps, with a different
sky coverage, and computing the estimate of $\tau$ in order to retrieve
the proper uncertainty scaling factor as function of $f_{\rm{sky}}$.}, we can
make a rough estimate of what our uncertainty should have been for
their analysis setup,
\begin{align}
  \sigma_{\mathrm{pred}} &\approx
  \sigma\cdot\sqrt{\frac{f_{\mathrm{sky}}^{\mathrm{BP}}}{f_{\mathrm{sky}}^{\mathrm{Natale}}}}\\\
  &= 0.013\cdot \sqrt{\frac{0.68}{0.54}}= 0.014.
\end{align}
For comparison, the actual \citet{natale:2020} uncertainty is 0.011, or about 30\,\%
smaller. We interpret our larger uncertainty as being due to marginalizing over a more
complete set of statistical uncertainties in the \BP\ analysis
framework than is possible with the frequentist-style and official LFI
and \WMAP\ data products. As such, this comparison directly highlights
the importance of the end-to-end approach.

\begin{figure}[t]
	\center
	\includegraphics[width=\linewidth]{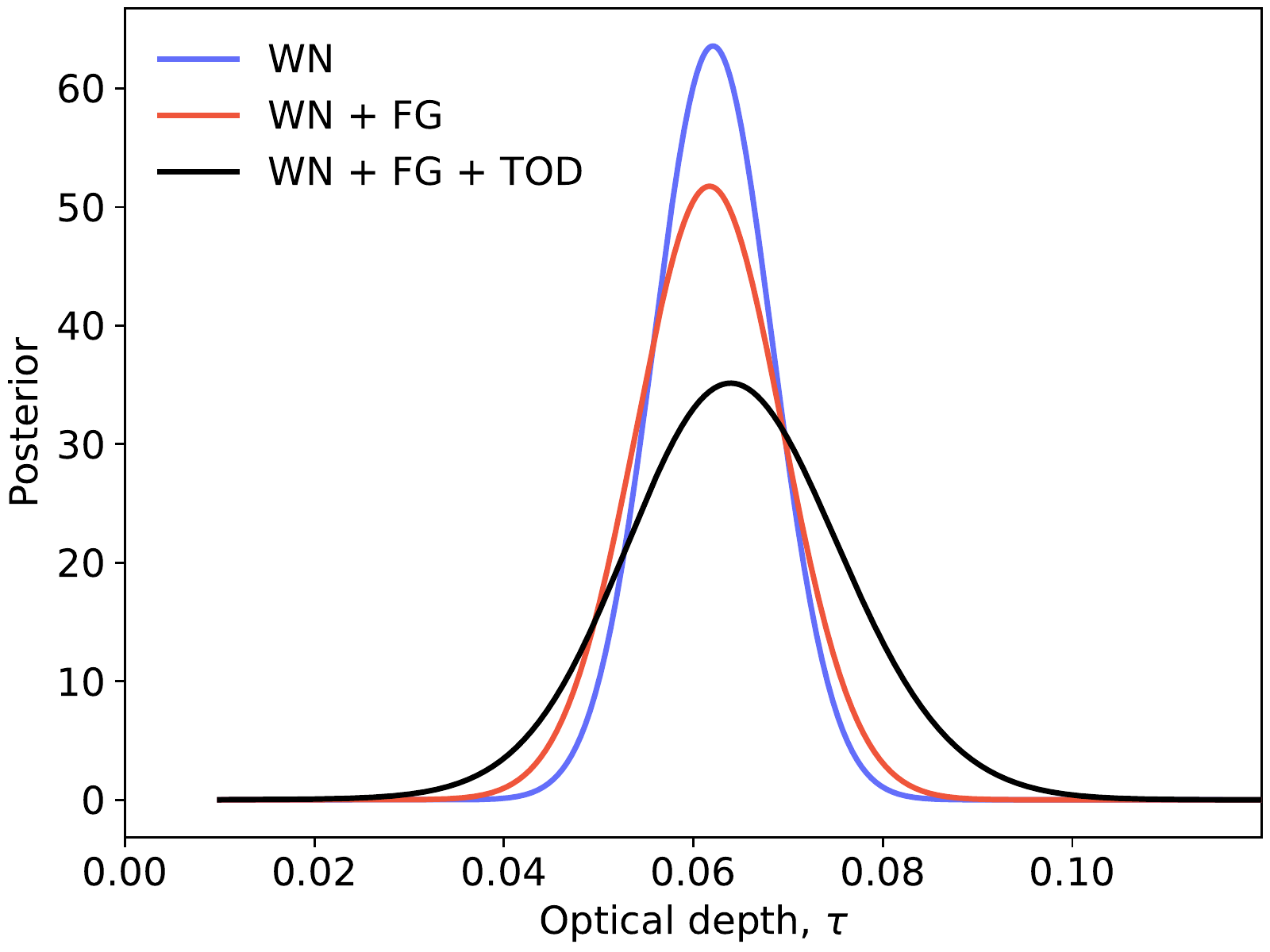}
        \caption{Estimates of $\tau$ under different uncertainty
          assumptions. The blue curve shows marginalization over white
          noise only; the red 
          curve shows marginalization over white noise and
          astrophysical uncertainties; and, finally, the black curve
          shows marginalization over all contributions, including
          low-level instrumental uncertainties, as in the
          final \BP\ analysis.}
	\label{fig:tau_assess}
\end{figure}

Table~\ref{tab:cospar_pol} also contains several goodness-of-fit and
stability tests. Specifically, we first note that the best-fit
tensor-to-scalar ratio is consistent with zero, and with an upper
95\,\% confidence limit of $r<0.6$. While this is by no means
competitive with current state-of-the-art constraints from the
combination of BICEP2/Keck and \Planck\ of $r<0.032$
\citep{tristram:2021}, the absence of strong $B$-mode power is a
confirmation that the \BP\ processing seems clean of systematic
errors; these results are in good agreement with the power spectrum
results presented by \citet{bp11}.

We also note in Table~\ref{tab:cospar_pol} that the impact of $\ell=2$
from the analysis is small, and the only noticeable effect of removing
it from the analysis is to increase the uncertainties on $\tau$ and
$r$ by about 10\,\%. This is important, because the \BP\ processing is
not guaranteed to have a unity transfer function for this single mode
($EE$, $\ell=2$): As discussed by \citet{bp07}, there is a strong
degeneracy between the CMB polarization quadrupole and the relative
gain parameters, and the current pipeline breaks this by imposing a
$\Lambda$CDM prior on the single $EE$ $\ell=2$ mode. Although this
effect is explicitly demonstrated through simulations to be small by
\citet{bp04}, it is still comforting to see that this particular mode
does not have a significant impact on the final results.

Finally, the sixth column in Table~\ref{tab:cospar_pol} shows the $\chi^2$
probability-to-exceed (PTE), where the main quantity is defined as
\begin{equation}
  \chi^2 = \hat{\s}_{\mathrm{CMB}}^t
  \left(\S(C_{\ell}^{\mathrm{bf}})+\N_{\mathrm{CMB}}\right)^{-1}\hat{\s}_{\mathrm{CMB}}.
\end{equation}
For a Gaussian and isotropic random field, this quantity should be
distributed according to a $\chi^2_{n_{\mathrm{dof}}}$ distribution,
where $n_{\mathrm{dof}}=225$ is the number of degrees of freedom,
which in our case is equal to the number of basis vectors in
$\hat{\s}_{\mathrm{CMB}}$. The PTE for our likelihood is 0.32,
indicating full consistency with the $\Lambda$CDM best-fit model\footnote{
In this paper we denote quantities fixed to a fiducial $\Lambda$CDM best-fit 
value with the superscript $\rm{bf}$.} 
and sample-based noise covariance matrix.\footnote{We note that this was
  not the case in the first preview version of the \BP\ results
  announced in November 2020: In that case the full-sky $\chi^2$ PTE
  was $\mathcal{O}(10^{-4})$, and this was eventually explained in
  terms of gain over-smoothing by \citet{bp07} and non-$1/f$
  correlated noise contributions by \citet{bp06}. Both these effects
  were mitigated in the final \BP\ processing, as reported here.}

Figure~\ref{fig:lowl_fsky} shows corresponding results for different
sky fractions, adopting the series of analysis masks defined by \citet{planck2016-l05}. The
tensor-to-scalar ratio is reported in terms of a nominal detection
level in units of $\sigma$, as defined by matching the observed
likelihood ratio $\mathcal{L}(r^{\mathrm{bf}})/\mathcal{L}(r=0)$ with
that of a Gaussian standard distribution. Overall, we see that all
results are largely insensitive to sky fraction, which suggests that
the current processing has managed to remove most statistically
significant astrophysical contamination \citep{bp13,bp14}. However, we do
note that a small $B$-mode contribution appears at the most aggressive
sky coverage of 83\,\%, and also that the $\chi^2$ PTE starts to fall
somewhat above 68\,\%. For this reason, we conservatively adopt a sky
fraction of 68\,\% for our main results, but note that 73\,\% would
have been equally well justified. 

Before concluding this section, we return to the importance of
end-to-end error propagation, and perform a simple analysis in which
we estimate the marginal $\tau$ posterior under three different regimes
of systematic error propagation. In the first regime, we
assume that the derived CMB sky map is entirely free of both
astrophysical and instrumental uncertainties, and the only source of
uncertainty is white noise. This case is evaluated by selecting one
random CMB sky map sample as the fiducial sky, and we do not
marginalize over instrumental or astrophysical samples when evaluating
the sky map and noise covariance matrix in Eq.~\eqref{eq:postmean}. In the second
regime, we assume that the instrumental model is perfectly known,
while the astrophysical model is uncertain. In the third and final
regime, we assume that both the instrumental and astrophysical
parameters are uncertain, and marginalize over everything, as in the
main \BP\ analysis. The results from these calculations are summarized
in Fig.~\ref{fig:tau_assess}. As expected, we see that the
uncertainties increase when marginalizing over additional
parameters. Specifically, the uncertainty of the fully marginalized
case is 46\,\% larger than for white noise, and 32\,\% larger than the
case marginalizing over the full astrophysical model. This calculation
further emphasizes the importance of global end-to-end analysis that
takes jointly into account all sources of uncertainty.

\section{Monte Carlo convergence}
\label{sec:convergence}

\begin{figure}[t]
  \center
  \includegraphics[width=\linewidth]{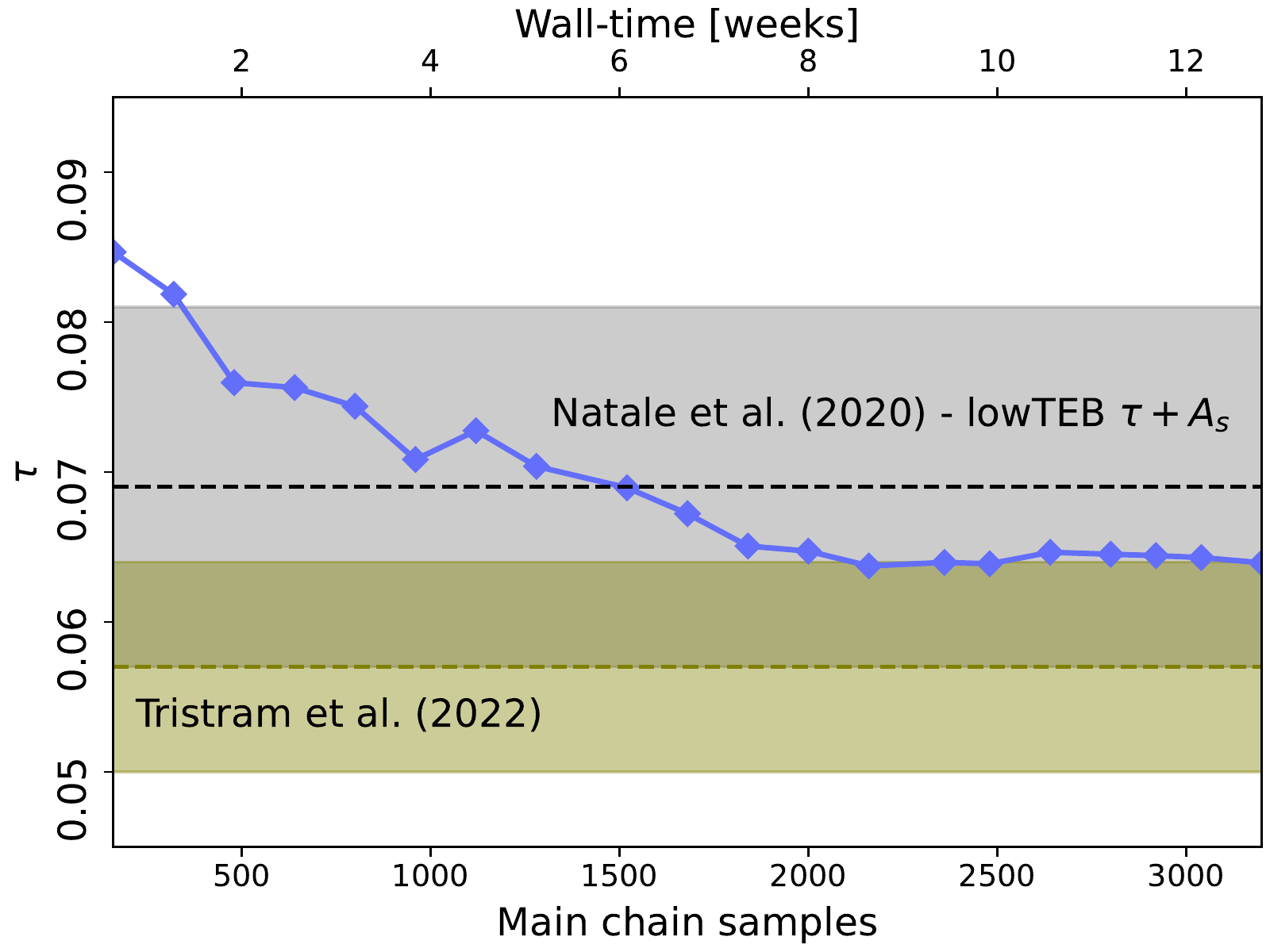}
  \caption{Convergence of constraints of the reionization optical depth as 
    a function of the number of main chain samples used to
    construct the CMB mean map and covariance matrix and the relative wall time needed
    to produce such samples in the main Gibbs loop. 
    The solid blue line shows the posterior mean for $\tau$, while the gray and green 
    regions show the corresponding 68\,\% confidence interval for \citet{natale:2020} and \citet{tristram:2021} respectively.}
    \label{fig:convergence}
\end{figure}

\begin{figure}[t]
  \center
  \includegraphics[width=\linewidth]{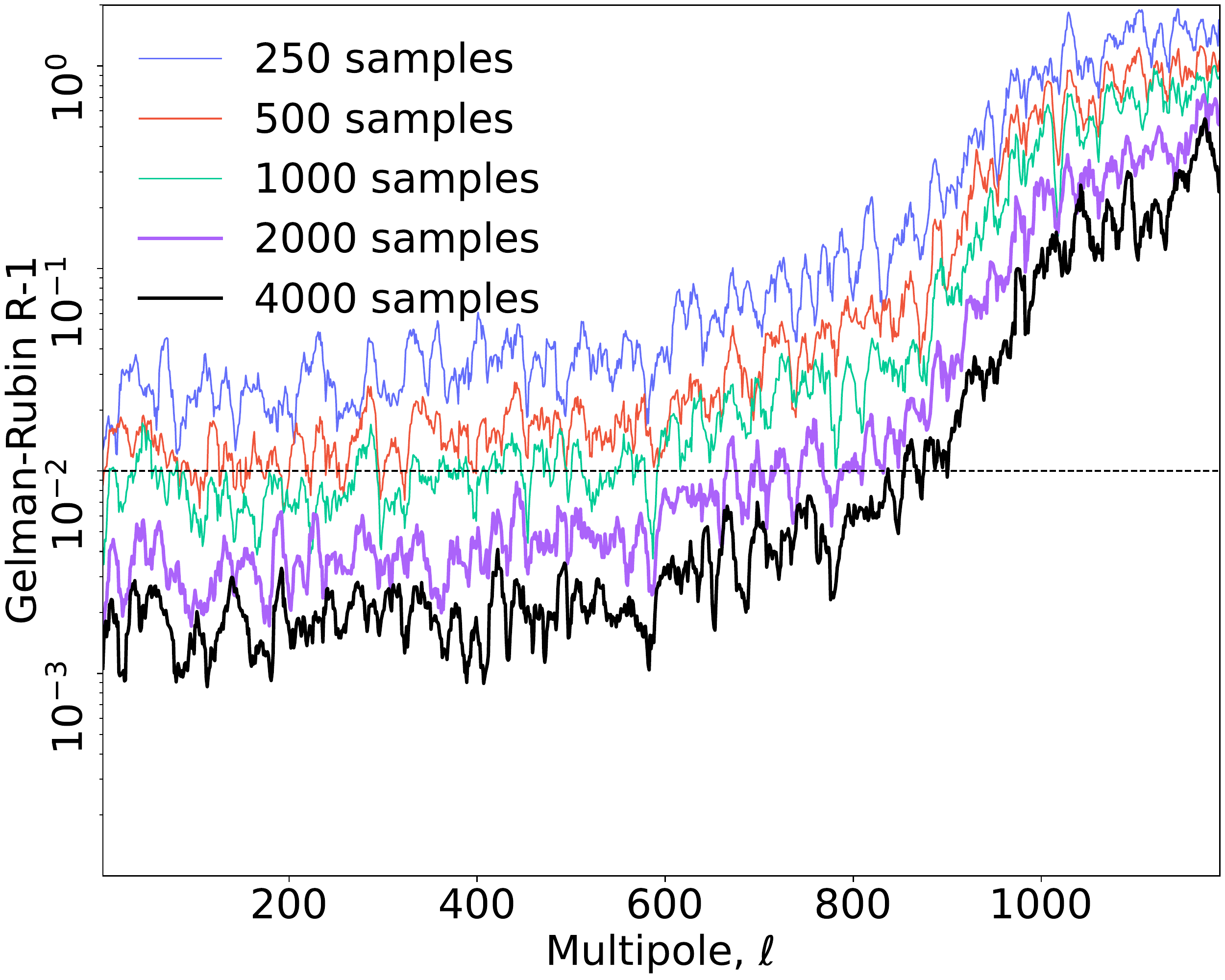}
  \caption{Gelman-Rubin convergence statistic for the \BP\ $\mathrm{TT}$
    angular power spectrum, as evaluated from four independent
    $\sigma_{\ell}$ chains. A $R-1$ value lower than 0.1 
    typically indicates
    acceptable convergence. Moreover, we report the $R-1=10^{-2}$ 
    threshold (dotted black line)
    representing a safer criterion to assess convergency.}
    \label{fig:gr_TT}
\end{figure}

As noted in Sect.~\ref{sec:bp}, one important goal of the current
paper is to assess how many end-to-end Monte Carlo samples are
required to robustly derive covariance matrices and cosmological
parameters by Gibbs sampling. We are now finally in a position to
answer this question quantitatively, using the results already
presented.

Starting with the low-$\ell$ polarization likelihood, we once again
adopt $\tau$ as a proxy for overall stability, and show in
Fig.~\ref{fig:convergence} $\tau$ as function of the number of Gibbs
samples, $n_{\mathrm{samp}}$, used to build the low-$\ell$ likelihood
inputs in Eq.~\ref{eq:postmean}.\footnote{Recall that for each main
  Gibbs chain sample, we additionally draw $n=50$ sub-samples to
  cheaply marginalize over white noise, such that the actual number of
  individual samples involved in Fig.~\ref{fig:convergence} is
  actually 50 times higher than what is shown; the important question
  for this test, however, is the number of main Gibbs samples.} Here
we see that the estimates are positively biased for small values of
$n_{\mathrm{samp}}$, with a central value around
$\tau=0.085$. However, the estimates then starts to gradually fall
while the Markov chains explore the full distribution. This behaviour
can be qualitatively understood as follows: The actual posterior mean
sky map converges quite quickly with number of samples, and stabilizes
only with a few hundred samples. However, the $\tau$ estimate is
derived by comparing the covariance of this sky map with the predicted
\emph{noise covariance} as given by $\N$; any excess fluctuations in
$\bar{\s}$ compared to $\N$ is interpreted as a positive $\S$
contribution. Convergence in $\N$ is obviously much more expensive 
than convergence in $\bar{\s}$, which leads to the slow decrease in 
$\tau$ as a function of sample as $\N$ becomes better 
described by more samples.

From Fig.~\ref{fig:convergence}, we see that the results stabilize
only after $n_{\mathrm{samp}}\approx 2000$ main Gibbs samples, which
is almost nine times more than the number of modes in the covariance
matrix, $n_{\mathrm{mode}}=225$. Obviously, this number will depend on
the specifics of the data models and datasets in question, and more
degenerate models will in general require more samples, but at least
this estimate provides a real-world number that may serve as a
rule-of-thumb for future analyses.

Finally, to assess convergence for the high-$\ell$ temperature
likelihood, we adopt the Gelman-Rubin (GR) $R$ convergence statistic,
which is defined as the ratio of the ``between-chain variance'' and
the ``in-chain variance'' \citep{gelman:1992}. We evaluate this
quantity based on the four available $\sigma_{\ell}$ chains, including
different numbers of samples in each case, ranging between 250 to
4000. The results from this calculations are summarized in
Fig.~\ref{fig:gr_TT}. Here we see that the convergence improves
rapidly below $\ell\lesssim600$--800, while multipoles above
$\ell\gtrsim 1000$ converge very slowly. We adopt a stringent
criterion of $R-1 < 0.01$ (dashed horizontal line), and conservatively
restrict the multipole range used by \BP\ to $\ell\le600$. With these
restrictions, we once again see that about 2000 samples are required
to converge.

\section{Conclusions}
\label{sec:conclusions}

The main motivation behind the \BP\ project is to develop a fully
Bayesian framework for global analysis of CMB and related datasets
that allows for joint analysis of both astrophysical and instrumental
effects, and thereby robust end-to-end error propagation. In this
paper, we have demonstrated this framework in terms of standard
cosmological parameters, which arguably represent the most valuable
deliverable for any CMB experiment. We emphasize that this work is
primarily algorithmic in nature, and intended to demonstrate the
Bayesian framework itself using a well-controlled dataset, namely the
\Planck\ LFI measurements; it is not intended to replace the current
state-of-the-art \Planck\ 2018 results, which are based on
high-sensitivity HFI measurements.

With this observation in mind, we find that the cosmological
parameters derived from LFI and \WMAP\ in \BP\ are overall in good
agreement with those published from the previous pipelines. When
considering the basic \LCDM\ parameters and temperature information
between $30\le\ell\le600$, the typical agreement between the various
cases is better than $1\,\sigma$, and we also note that in the cases
where there are discrepancies, the \BP\ results are typically somewhat
closer to the high-$\ell$ HFI constraints than previous results,
indicating less internal tension between low and high multipoles.

Overall, the most noticeable difference is seen for the optical depth
of reionization, for which we find a slightly higher value of
$\tau=0.066\pm0.013$ than \Planck\ 2018 at $\tau=0.051\pm0.006$. At
the same time, this value is lower than the corresponding
LFI-plus-\WMAP\ result derived by \citet{natale:2020} of
$\tau=0.069\pm0.011$, which suggests that the current processing has
cleaned up more systematic errors than in previous LFI
processing. Furthermore, and even more critically, we find that the
\BP\ uncertainty is almost 30\,\% larger than latter when taking into
account the different sky fraction, and we argue that this is
due to \BP\ taking into account a much richer systematic error model
than previous pipelines. Indeed, this result summarizes the main
purpose of the entire \BP\ project in terms of one single number. We
believe that this type of global end-to-end processing will be
critical for future analysis of next-generation $B$-mode experiments.

A second important goal of the current paper was to quantify how many
samples are actually required to converge for a Monte Carlo-based
approach. Based on the current analysis, we find that about 2000
end-to-end samples are need to achieve robust results. Obviously,
introducing additional sampling steps that more efficiently break down
long Markov chain correlation lengths will be important to reduce this
number in the future, but already the current results proves that the
Bayesian approach is computationally feasible for past and current
experiments. 

\begin{acknowledgements}
  We thank Prof.\ Pedro Ferreira and Dr.\ Charles Lawrence for useful suggestions, comments and 
  discussions. We also thank the entire \Planck\ and \WMAP\ teams for
  invaluable support and discussions, and for their dedicated efforts
  through several decades without which this work would not be
  possible. The current work has received funding from the European
  Union’s Horizon 2020 research and innovation programme under grant
  agreement numbers 776282 (COMPET-4; \BP), 772253 (ERC;
  \textsc{bits2cosmology}), and 819478 (ERC; \textsc{Cosmoglobe}). In
  addition, the collaboration acknowledges support from ESA; ASI and
  INAF (Italy); NASA and DoE (USA); Tekes, Academy of Finland (grant
   no.\ 295113), CSC, and Magnus Ehrnrooth foundation (Finland); RCN
  (Norway; grant nos.\ 263011, 274990); and PRACE (EU).
\end{acknowledgements}

\bibliography{Planck_bib,BP_bibliography,BP_cosmoparams_BIB}
\bibliographystyle{aa}

\raggedright
\end{document}